\documentclass[journal]{IEEEtran}

\usepackage{graphicx}

\newtheorem{theorem}{\bf Theorem}

\newtheorem{condition}{\bf Condition}

\newtheorem{example}{\bf Example}

\begin{document}
%
\title{Distributed Multiple Access with Multiple Transmission Options at The Link Layer}
%
%
\author{Faeze Heydaryan, Yanru Tang, and Jie Luo
\thanks{The authors are with the Electrical and Computer Engineering Department, Colorado State University, Fort Collins, CO 80523. E-mail: \{faezeh66, yrtang, rockey\}@colostate.edu. }
\thanks{This work was supported by the National Science Foundation under Grants CCF-1420608 and CNS-1618960. Any opinions, findings, and conclusions or recommendations expressed in this paper are those of the authors and do not necessarily reflect the views of the National Science Foundation.}
}

\maketitle

\begin{abstract}
This paper investigates the problem of distributed medium access control in a wireless multiple access network with an unknown finite number of homogeneous transmitters. An enhanced physical link layer interface is considered where each link layer user can be equipped with multiple transmission options. Assume that each user is backlogged with a saturated message queue. With a generally-modeled channel, a distributed medium access control framework is suggested to adapt the transmission scheme of each user to maximize an arbitrarily chosen symmetric network utility. The proposed framework suggests that the receiver should measure the success probability of a carefully designed virtual packet, and feed such information back to the transmitters. Upon receiving the measured probability, each transmitter should obtain a user number estimate by comparing the probability with its theoretical value, and then adapt its transmission scheme accordingly. Conditions under which the proposed algorithm should converge to a designed unique equilibrium are characterized. Simulation results are provided to demonstrate the optimality and the convergence properties of the proposed algorithm.
\end{abstract}

\begin{keywords}
medium access control, distributed system, adaptive algorithms, wireless network, multiple access communication
\end{keywords}

%
\IEEEpeerreviewmaketitle


\section{Introduction}
\label{S-Introduction}
Due to the increasing dynamics of communication activities, a significant proportion of messages in communication networks are transmitted using distributed protocols where users make their transmission decisions and communication parameter choices individually. Classical network architecture such as the OSI model assumes that each link layer user should be equipped with a single transmission option plus an idling option \cite{ref Bertsekas92}. At any moment, a link layer user can only choose to idle or to transmit a packet. When communication cannot be fully optimized at the physical layer, which happens often in a distributed wireless network, data link layer must share the responsibility of transmission adaptation. However, the single transmission option setting significantly limited the capability of exploiting advanced wireless tools such as rate and power adaptations at the data link layer.

Recently, a new channel coding theory was proposed in \cite{ref Luo12}\cite{ref Wang12}\cite{ref Luo15} for distributed communication at the physical layer. The coding theory allows each physical layer transmitter to prepare an ensemble of channel codes, and to choose an arbitrary one (according to the link layer decision) to encode its message and to transmit the codeword symbols to the receiver. While code ensembles of the users are assumed to be known, actual coding decisions are not shared among the transmitters or with the receiver. The receiver, on the other hand, should either decode the messages of interest or report collision, depending on whether a pre-determined error probability requirement can be met. Fundamental limit of the system was characterized using an achievable region defined in the vector space of the coding decisions of the transmitters. The achievable region was shown in \cite{ref Luo12} to coincide with the classical Shannon region. Error performance bounds in the case of finite codeword length were obtained in \cite{ref Wang12}\cite{ref Luo15}.

The new channel coding theory provided the basic physical layer support for an enhancement to the physical-link layer interface \cite{ref Luo15}, which allows each link layer user to be equipped with multiple transmission options. These options correspond to different codes at the physical layer, possibly representing different communication settings such as different transmission power and rate combinations. The interface enhancement enables data link layer protocols to exploit advanced wireless communication adaptations through the navigation of different transmission options. This is a much needed capability for mitigating architectural inefficiency at the bottom two layers of many wireless networks. However, to maintain a layered network architecture (or system modularity), a link layer user is constrained to the provided options for transmission adaptation. How should a user efficiently exploit the often limited options to optimize a network utility, is a key question that needs to be answered.

Distributed medium access control (MAC) protocols can be categorized into non-adaptive ALOHA protocols \cite{ref Abramson70}, splitting algorithms \cite{ref Gallager78}, and back-off approaches \cite{ref Hajek82}\cite{ref Hajek85}\cite{ref Bianchi00}. ALOHA protocols have been widely used to investigate fundamental network properties, such as achievable throughput and stability regions \cite{ref Luo06}. In splitting algorithms such as the FCFS algorithm \cite{ref Gallager78}, each user maintains a common virtual interval and a randomly generated identity value belonging to the interval. Users partition the interval and order the sub-intervals based upon a sequence of channel feedback messages. Transmission schedule of the users are determined accordingly. While splitting algorithms can often achieve a relatively high system throughput, their correct function depends on the assumptions of instant availability of noiseless channel feedback and correct reception of feedback sequence. Both of these conditions, unfortunately, can be violated in a wireless environment. Theoretical analysis of a splitting algorithm can be extremely challenging, especially when wireless-related factors such as channel noise, feedback error, and transmission delay are taken into account. Back-off algorithms, on the other hand, has proven to enjoy more trackable analysis \cite{ref Hajek82}\cite{ref Hajek85}\cite{ref Bianchi00}. In back-off algorithms such as the 802.11 DCF protocol, depending on packet availability, each user transmits its packets randomly according to an associated probability parameter. A user should decrease its transmission probability in response to a packet collision (or transmission failure) event, and increase its transmission probability in response to a transmission success event. Distributed probability adaptation in a back-off algorithm often falls into the framework of stochastic approximation algorithms \cite{ref Hajek82}\cite{ref Hajek85}, with rigorously developed mathematical and statistical tools available for its performance analysis. It is well known that convergence proof of these algorithms often hold in the existence of measurement noise and feedback delay \cite{ref Kushner97}. Practical back-off algorithms can also be analyzed using Markov models to characterize the impact of discrete probability updates \cite{ref Bianchi00}.

In \cite{ref Hajek85}, a stochastic approximation model was proposed for distributed networking over a collision channel with an unknown finite number of users, each having a saturated message queue. By targeting the transmission probability of each user as a function of a locally measurable system variable, such as the channel idling probability, it was shown that the system can be designed to converge to a unique stable equilibrium. In the case of throughput maximization with homogeneous users, it was proposed that idling probability of the channel should be controlled toward the asymptotically optimal value of $1/e$. This is similar to the proposal of controlling the total traffic level toward $1$, as discussed in \cite{ref Hajek82} using a stochastic approximation framework for a system with an infinite number of users. Most of the existing analysis of the splitting and the back-off algorithms either assumes a throughput optimization objective and/or a simple collision channel model. While significant research efforts have been made to revise collision resolution algorithms to incorporate wireless-related physical layer properties, such as capture effect \cite{ref Lau92} and multi-packet reception \cite{ref Ghez88}, not much progress has been reported since the 1980s on integrating these extensions with the insightful stochastic approximation-based frameworks, such as those introduced in \cite{ref Hajek82}\cite{ref Hajek85}.

With the enhanced physical-link layer interface, a link layer user can be equipped with multiple transmission options. Link layer networking can face a set of channel models that is much richer and more complicated than the classical collision channel. It is not immediately clear how collision resolution should be done in such a scenario. For example, if a user can adapt its transmission power and rate in addition to its transmission probability, what does ``back-off'' even mean in this case? Motivated by this and similar simple questions, in this paper, we investigate the problem of distributed MAC in a wireless multiple access network with/without the enhanced physical-link layer interface. To maintain a relatively simple and trackable investigation, we assume that the network should have an unknown finite number of homogeneous users (transmitters), each being backlogged with a saturated message queue. Other than the user homogeneity assumption, our choice of problem formulation and analytical tools are similar to those presented in \cite{ref Hajek85} for the collision channel. First, the assumption of saturated message queues is introduced to avoid the complication of random message arrivals. While bursty message arrival is rather an important character of distributed network systems \cite{ref Bertsekas92}\cite{ref Ephremides98}, it is known to create coupling between transmission activities of the users. Such coupling often leads to open research problems in throughput and stability analysis of systems with a relatively small number of users \cite{ref Rao88}\cite{ref Luo06}. Results obtained with the assumption of saturated message queues can often serve as achievable bounds to the corresponding results for systems with random message arrivals. Second, because each user only interacts with the receiver, the assumption of multiple access networking with homogeneous users mainly represents the communication environment envisioned by each link layer user. In other words, without further knowledge about the actual networking environment, a link layer protocol should be designed to help a user to get a fair share of the multiple access channel under the assumption of user homogeneity\footnote{Note that user symmetry is widely assumed in many channel models such as the collision channel \cite{ref Bertsekas92} and the multi-packet reception channel \cite{ref Ghez88}.}. While it is possible to extend the system model to the case of heterogeneous users by following the approach presented in \cite{ref Hajek85}, such an extension and its analysis are outside the scope of this paper. Third, because users in a distributed network often access the channel opportunistically, it may not be easy to know how many users are actually active. We assume that each user should be able to calculate its optimal transmission scheme if the user number is known, but we would like to develop distributed algorithms to lead the system to a close-to-optimal operation point without the knowledge of the actual user number. The expectation is that, if fast adaptation algorithms can be developed accordingly, a system can keep track of the active user number even if users frequently join/exist the communication party.

The rest of the paper is organized as follows. In Section \ref{S-StochasticApproximation}, we present a stochastic approximation framework for a class of distributed MAC algorithms with guaranteed convergence to a unique system equilibrium. While the results are more or less standard in the stochastic approximation literature, they characterize the key conditions for convergence. Within the framework, the research problem becomes how one should design the system to place the unique equilibrium at the desired point that maximizes a chosen network utility and to make sure the conditions for convergence are satisfied. In Section \ref{SingleOption}, we consider the case when each user has a single transmission option. A distributed MAC algorithm is proposed to adapt the transmission probability of each user according to a channel contention measure, which is defined as the success probability of a virtual packet. The MAC algorithm is then extended in Section \ref{MultipleOptions} to the case when users have multiple transmission options. Simulation results are provided in Section \ref{Simulation} to demonstrate both the optimality and the convergence properties of the proposed MAC algorithms.

\section{A Stochastic Approximation Framework}
\label{S-StochasticApproximation}
Consider a distributed multiple access network with a memoryless channel and $K$ homogeneous users (transmitters). Time is slotted. The length of each time slot equals the transmission duration of one packet. We assume that user number $K$ should be unknown to the users and also unknown to the receiver. Each user is equipped with $M$ transmission options plus an idling option, and is backlogged with a saturated message queue. At the beginning of each time slot $t$, a user should either idle or randomly choose a transmission option to send a message, with corresponding probabilities being specified by an associated probability vector. Transmission decisions of the users are made individually, and they are shared neither among the users nor with the receiver. The $M$-length probability vector associated to user $k$, $k=1, \dots, K$, is denoted by $\mbox{\boldmath $p$}_k (t)$ for time slot $t$. We write $\mbox{\boldmath $p$}_k (t)=p_k(t) \mbox{\boldmath $d$}_k (t)$, with $0\le p_k(t) \le 1$ being the probability that user $k$ transmits a packet in time slot $t$, and with vector $\mbox{\boldmath $d$}_k (t)$ specifying the conditional probabilities for user $k$ to choose each of the transmission options should it decide to transmit a packet. Entries of the $\mbox{\boldmath $d$}_k (t)$ vector satisfy $0\le d_{km}(t)\le 1$ for $1\le m \le M$, and $\sum_{m=1}^M d_{km}(t)=1$. We term $p_k(t)$ the ``transmission probability'' of user $k$, and term $\mbox{\boldmath $d$}_k (t)$ the ``transmission direction'' vector of user $k$.

At the end of each time slot $t$, based upon available channel feedback, each user $k$ derives a target probability vector $\mbox{\boldmath $\tilde{p}$}_k (t)$. User $k$ then updates its transmission probability vector by
\begin{eqnarray}
&& \mbox{\boldmath $p$}_k (t+1)=(1-\alpha(t))\mbox{\boldmath $p$}_k (t)+\alpha(t)\tilde{\mbox{\boldmath $p$}}_k(t) \nonumber \\
&& =\mbox{\boldmath $p$}_k (t)+\alpha(t)(\tilde{\mbox{\boldmath $p$}}_k(t)-\mbox{\boldmath $p$}_k(t)),
\label{UserProbabilityVectorUpdate}
\end{eqnarray}
where $\alpha(t)>0$ is a step size parameter of time slot $t$. Let $\mbox{\boldmath $P$}(t)=[\mbox{\boldmath $p$}_1^T (t), \mbox{\boldmath $p$}_2^T (t), \dots, \mbox{\boldmath $p$}_K^T (t) ]^T$ denote an $MK$-length vector that consists of the transmission probability vectors of all users in time slot $t$. Let $\tilde{\mbox{\boldmath $P$}}(t)=[\tilde{\mbox{\boldmath $p$}}_1^T (t), \tilde{\mbox{\boldmath $p$}}_2^T (t), \dots, \tilde{\mbox{\boldmath $p$}}_K^T (t) ]^T$ denote the corresponding target vector. According to (\ref{UserProbabilityVectorUpdate}), $\mbox{\boldmath $P$}(t)$ is updated by
\begin{equation}
\mbox{\boldmath $P$}(t+1)=\mbox{\boldmath $P$}(t)+\alpha(t)(\tilde{\mbox{\boldmath $P$}}(t)-\mbox{\boldmath $P$}(t)).
\label{StochasticProbabilityVectorUpdate}
\end{equation}
Probability adaptation given in (\ref{StochasticProbabilityVectorUpdate}) falls into the stochastic approximation framework \cite{ref Kushner97}\cite{ref Karlin75}\cite{ref Borkar00}, where the target probability vector $\tilde{\mbox{\boldmath $P$}}(t)$ is often calculated from noisy estimates of certain system variables, e.g., the channel idling probability.

Define $\hat{\mbox{\boldmath $P$}}(t)=[\hat{\mbox{\boldmath $p$}}_1^T (t), \hat{\mbox{\boldmath $p$}}_2^T (t), \dots, \hat{\mbox{\boldmath $p$}}_K^T (t) ]^T$ as the ``theoretical value'' of $\tilde{\mbox{\boldmath $P$}}(t)$ under the assumption that there is no measurement noise and no feedback error in time slot $t$. Let $E_t[\tilde{\mbox{\boldmath $P$}}(t)]$ be the conditional expectation of $\tilde{\mbox{\boldmath $P$}}(t)$ given system state at the beginning of time slot $t$. The difference between $E_t[\tilde{\mbox{\boldmath $P$}}(t)]$ and $\hat{\mbox{\boldmath $P$}}(t)$ is defined as the bias in the target probability vector calculation, denoted by $\mbox{\boldmath $G$}(t)$.
\begin{equation}
\mbox{\boldmath $G$}(t)=E_t[\tilde{\mbox{\boldmath $P$}}(t)]-\hat{\mbox{\boldmath $P$}}(t).
\end{equation}
Note that, given the communication channel, both $\hat{\mbox{\boldmath $P$}}(t)= \hat{\mbox{\boldmath $P$}}(\mbox{\boldmath $P$}(t))$ and $\mbox{\boldmath $G$}(t)=\mbox{\boldmath $G$}(\mbox{\boldmath $P$}(t))$ are functions of $\mbox{\boldmath $P$}(t)$, which is the transmission probability vector in time slot $t$.

The following two conditions are typically required for the convergence of a stochastic approximation algorithm \cite{ref Kushner97}\cite{ref Karlin75}\cite{ref Borkar00}.

\begin{condition}{\label{MeanBiasAssumption}}
(Mean and Bias) There exists a constant $K_m>0$ and a bounding sequence $0\le \beta(t) \le 1$, such that
\begin{equation}
\|\mbox{\boldmath $G$}(\mbox{\boldmath $P$}(t))\| \le K_m \beta(t).
\end{equation}
We assume that $\beta(t)$ is controllable in the sense that one can design protocols to ensure $\beta(t)\le \epsilon$ for any chosen $\epsilon>0$ and for large enough $t$.
\end{condition}

\begin{condition} {\label{LipschitzAssumption}}
(Lipschitz Continuity) There exists a constant $K_l>0$, such that
\begin{equation}
\|\hat{\mbox{\boldmath $P$}}(\mbox{\boldmath $P$}_a)-\hat{\mbox{\boldmath $P$}}(\mbox{\boldmath $P$}_b)\| \le K_l \|\mbox{\boldmath $P$}_a-\mbox{\boldmath $P$}_b\|, \mbox{ for all } \mbox{\boldmath $P$}_a, \mbox{\boldmath $P$}_b.
\end{equation}
\end{condition}

According to stochastic approximation theory \cite{ref Kushner97}, if the above two conditions are satisfied, the step size sequence $\alpha(t)$ and the bounding sequence $\beta(t)$ are small enough, then trajectory of the transmission probability vector $\mbox{\boldmath $P$}(t)$ under distributed adaptation given in (\ref{StochasticProbabilityVectorUpdate}) can be approximated by the following associated ordinary differential equation (ODE),
\begin{equation}
\frac{d \mbox{\boldmath $P$}(t)}{dt}=-[\mbox{\boldmath $P$}(t)-\hat{\mbox{\boldmath $P$}}(t)],
\label{AssociatedSystemODE}
\end{equation}
where we used $t$ to denote the continuous time variable. Because all entries of $\mbox{\boldmath $P$}(t)$ and $\hat{\mbox{\boldmath $P$}}(t)$ stay in the range of $[0, 1]$, any equilibrium $\mbox{\boldmath $P$}^* = [\mbox{\boldmath $p$}_1^{*T}, \dots, \mbox{\boldmath $p$}_K^{*T}]^T$ of the associated ODE must satisfy
\begin{equation}
\mbox{\boldmath $P$}^*=\hat{\mbox{\boldmath $P$}}(\mbox{\boldmath $P$}^*).
\label{SystemEquilibriumSolution}
\end{equation}

Suppose that the associated ODE given in (\ref{SystemEquilibriumSolution}) has a unique solution at $\mbox{\boldmath $P$}^*$, then the following convergence results can be obtained from the standard conclusions in the stochastic approximation literature.

\begin{theorem}{\label{StochasticProbabilityOneConvergence}}
For distributed transmission probability adaptation given in (\ref{StochasticProbabilityVectorUpdate}), assume that the associated ODE given in (\ref{AssociatedSystemODE}) has a unique stable equilibrium at $\mbox{\boldmath $P$}^*$. Suppose that $\alpha(t)$ and $\beta(t)$ satisfy the following conditions
\begin{equation}
\sum_{t=0}^{\infty}\alpha(t)=\infty, \sum_{t=0}^{\infty}\alpha(t)^2 <\infty, \sum_{t=0}^{\infty}\alpha(t)\beta(t)<\infty.
\label{ProbabilityOneSequnceRequirement}
\end{equation}
Under Conditions \ref{MeanBiasAssumption} and \ref{LipschitzAssumption}, $\mbox{\boldmath $P$}(t)$ converges to $\mbox{\boldmath $P$}^*$ with probability one.
\end{theorem}

Theorem \ref{StochasticProbabilityOneConvergence} is implied by \cite[Theorem 4.3]{ref Kushner97}.

\begin{theorem}{\label{StochasticWeakConvergence}}
For distributed transmission probability adaptation given in (\ref{StochasticProbabilityVectorUpdate}), assume that the associated ODE given in (\ref{AssociatedSystemODE}) has a unique stable equilibrium at $\mbox{\boldmath $P$}^*$. Let Conditions \ref{MeanBiasAssumption} and \ref{LipschitzAssumption} hold true. Then for any $\epsilon>0$, there exists a constant $K_w>0$, such that, for any $0<\underline{\alpha}<\overline{\alpha}<1$ satisfying the following constraint
\begin{equation}
\exists T_0\ge 0, \underline{\alpha}\le \alpha(t) \le \overline{\alpha}, \beta(t)\le \sqrt{\overline{\alpha}}, \forall t\ge T_0,
\label{ProbabilitySequnceRequirement}
\end{equation}
$\mbox{\boldmath $P$}(t)$ converges weakly to $\mbox{\boldmath $P$}^*$ in the following sense
\begin{equation}
\mathop{\lim\sup}_{t\to \infty} Pr\left\{\| \mbox{\boldmath $P$}(t)- \mbox{\boldmath $P$}^* \| \ge \epsilon \right\} < K_w \overline{\alpha}.
\end{equation}
\end{theorem}

Theorem \ref{StochasticWeakConvergence} can be obtained by following the proof of \cite[Theorem 2.3]{ref Borkar00} with minor revisions.

For simplicity, we assumed the same step size sequence $\alpha(t)$ and the same bounding sequence $\beta(t)$ for all users. We also assumed that all users should update their transmission probability vectors synchronously in each time slot. However, by following the literature of stochastic approximation theory \cite{ref Kushner97}, it is easy to show that different users can use different step sizes and bounding sequences, and can also adapt their probability vectors asynchronously. Convergence results stated in Theorems \ref{StochasticProbabilityOneConvergence} and \ref{StochasticWeakConvergence} should remain valid, if the step sizes and bounding sequences of all users satisfy the same constraints given in (\ref{ProbabilityOneSequnceRequirement}) and (\ref{ProbabilitySequnceRequirement}), and users also update their probability vectors frequently enough.

Theorems \ref{StochasticProbabilityOneConvergence} and \ref{StochasticWeakConvergence} provided convergence guarantee for a class of distribute MAC algorithms. Within the presented stochastic approximation framework, the key question is how to design a distributed MAC algorithm to satisfy Conditions \ref{MeanBiasAssumption} and \ref{LipschitzAssumption} and to place the unique equilibrium of the associated ODE at a point that maximizes a chosen utility. Because users are homogeneous, if equilibrium of the system is unique, transmission probability vectors of the users at the equilibrium must be identical. We choose to enforce such a property by guaranteeing that all users should obtain the same target transmission probability vector in each time slot. This is achieved by the following design details.

We assume that, in each time slot, there is a virtual packet being transmitted through the channel. Virtual packets assumed in different time slots are identical. A virtual packet is an assumed packet whose coding parameters are known to the users and to the receiver, but it is not physically transmitted in the system, i.e., the packet is ``virtual''. Without knowing the transmission/idling status of the users, we assume that the receiver can detect whether the reception of a virtual packet should be regarded as successful or not, and therefore can estimate its success probability. For example, suppose that the link layer channel is a collision channel, and a virtual packet has the same coding parameters of a real packet. Then, virtual packet reception in a time slot should be regarded as successful if and only if no real packet is transmitted. Success probability of the virtual packet in this case equals the idling probability of the collision channel. For another example, if all packets including the virtual packet are encoded using random block codes, given the physical layer channel, reception of the virtual packet corresponds to a detection task that judges whether or not the vector transmission status of all real users should belong to a specific region. Such detection tasks and their performance bounds have been extensively discussed in the distributed channel coding literature \cite{ref Luo12}\cite{ref Wang12}\cite{ref Luo15}.

Let $q_v(t)$ denote the success probability of the virtual packet in time slot $t$. We term $q_v(t)$ the ``channel contention measure'' because it is designed to serve as a measurement of the contention level of the link-layer
multiple access channel. We assume that the receiver should obtain an estimate of $q_v(t)$ and feed it back to all transmitters. Note that, in the collision channel case when $q_v(t)$ equals the channel idling probability, feeding back an estimate of $q_v(t)$ may not be necessary. So long as each user $k$ knows the success probability of its own packet, denoted by $q_k(t)$, idling probability of the channel can be calculated by $(1-p_k(t))q_k(t)$. With a general link layer channel, however, obtaining $q_v(t)$ is not always possible if an estimate is not fed back directly by the receiver. Upon receiving the estimate of $q_v(t)$, each user calculates its target probability vector as the same function of the $q_v(t)$ estimate. Denote the theoretical target probability vector of a user by $\mbox{\boldmath $\hat{p}$}(q_v(t))$. The theoretical target probability vector of all users is given by $\mbox{\boldmath $\hat{P}$}(t)=\mbox{\boldmath $1$}\otimes \mbox{\boldmath $\hat{p}$}(q_v(t))$, where $\mbox{\boldmath $1$}$ denotes a $K$-length vector of all $1$'s and $\otimes$ represents the Kronecker product. Consequently, according to (\ref{AssociatedSystemODE}), any equilibrium $\mbox{\boldmath $P$}^*$ of the ODE must take the form of $\mbox{\boldmath $P$}^*=\mbox{\boldmath $1$}\otimes \mbox{\boldmath $p$}^*$. Because $q_v$ is a function of the transmission probability vectors of all users, we must have $\mbox{\boldmath $P$}^*=\mbox{\boldmath $1$}\otimes \mbox{\boldmath $p$}^*=\mbox{\boldmath $1$}\otimes \mbox{\boldmath $\hat{p}$}(\mbox{\boldmath $p$}^*)$, where $\mbox{\boldmath $\hat{p}$}(\mbox{\boldmath $p$}^*)$ denotes the derived theoretical target probability vector of a user given that all users have the same transmission probability vector $\mbox{\boldmath $p$}^*$.

In a practical system, an estimate of $q_v(t)$ is likely to be corrupted by measurement noise. We assume that, if users keep their transmission probability vector $\mbox{\boldmath $P$}$ at a constant, and $q_v$ is measured over an interval of $Q$ time slots, then the measurement should converge to its true value with probability one as $Q$ is taken to infinity. Other than this assumption, measurement noise is not involved in the discussion of the design objectives, i.e., to meet Conditions \ref{MeanBiasAssumption} and \ref{LipschitzAssumption} and to place the unique system equilibrium at the desired point. Therefore, in the following two sections, we assume that $q_v(t)$ can be measured precisely and be fed back to the users. This leads to $\mbox{\boldmath $\tilde{P}$}(t)=\mbox{\boldmath $\hat{P}$}(t)= \mbox{\boldmath $1$}\otimes \mbox{\boldmath $\hat{p}$}(t)$. We will also skip time index $t$ to simplify the notations.

\section{Single Transmission Option}
\label{SingleOption}
Let us first consider the simple situation when each user is equipped with a single transmission option ($M=1$) plus an idling option. In this case, each user $k$ should maintain a scalar transmission probability parameter, denoted by $p_k$. Transmission probabilities of all users are listed in a $K$-length vector, denoted by $\mbox{\boldmath $p$} = [p_1, \dots, p_K]^T$. We will show that, with a general channel model and without knowing the user number $K$, a distributed MAC algorithm can be designed to lead the system to a unique equilibrium that is not far from optimal with respect to a chosen symmetric network utility.

Given the physical layer channel and the coding details of the packets, we model the data link layer multiple access channel using two sets of channel parameters. The first set is termed the ``real channel parameter set", denoted by $\{C_{rj}\}$ for $j \ge 0$. $C_{rj}$ is the conditional success probability of a real packet should it be transmitted in parallel with $j$ other real packets. The second set is termed the ``virtual channel parameter set'', denoted by $\{C_{vj}\}$ for $j \ge 0$. $C_{vj}$ is the success probability of the virtual packet should it be transmitted in parallel with $j$ real packets. We assume that $C_{vj} \ge C_{v(j+1)}$ should hold for all $j\ge 0$, which means that an increased number of parallel real packet transmissions should not help a virtual packet to get through the channel. Let $\epsilon_v>0$ be a pre-determined small constant. We define $J_{\epsilon_v}$ as the minimum integer such that $C_{v J_{\epsilon_v}}$ is strictly larger than $C_{v (J_{\epsilon_v}+1)}+\epsilon_v$, i.e.,
\begin{equation}
  J_{\epsilon_v}  =\mathop{\arg \min}_j C_{vj} > C_{v(j+1)} +\epsilon_v.
  \label{DefiningJ_epsilon_v_Single}
\end{equation}
Because $\{C_{rj}\}$ and $\{C_{vj}\}$ can be derived from the physical layer channel model and the coding details of the packets \cite{ref Luo12}\cite{ref Wang12}\cite{ref Luo15}, we assume that both of them should be known at the users and at the receiver. Note that, while virtual packet is irrelevant to the derivation of $\{C_{rj}\}$, its coding parameters do affect the values of $\{C_{vj}\}$.

We assume that users intend to maximize a symmetric network utility, denoted by $U(K, p, \{C_{rj}\})$. Under the assumption that all users should transmit with the same probability $p$, i.e., $\mbox{\boldmath $p$}=p\mbox{\boldmath $1$}$, system utility is a function of the user number $K$, the common transmission probability $p$, and the real channel parameter set $\{C_{rj}\}$. For example, if we choose the utility function as the sum system throughput, then $U(K, p, \{C_{rj}\})$ should be given by
\begin{eqnarray}
&& U(K, p, \{C_{rj} \}) = \nonumber \\
&& \quad K \sum_{j=0}^{K-1} \left( {K-1 \atop j} \right)  p^{j+1} (1-p)^{K-1-j} C_{rj}.
\label{ThroughputSingle}
\end{eqnarray}
For most of the utility functions of interest, such as the sum throughput function given above, an asymptotically optimal solution should keep the expected load of the channel at a constant \cite{ref Hajek85}\cite{ref Ghez88}. Therefore, if $p_K^*$ is the optimal transmission probability for user number $K$, we should have $\lim_{K \to \infty} K p_K^* = x^*$ with $x^*>0$ being obtained by the following asymptotic utility optimization.
\begin{equation}
 x^* = \mathop{\arg \max}_x \lim_{K \to \infty} U\left(K, \frac{x}{K}, \{C_{rj}\}\right).
 \label{xstar}
\end{equation}

Without knowing the actual user number $K$, we will show next that it is possible to set the system equilibrium at $\mbox{\boldmath $p$}^*=\min \left\{p_{\max}, \frac{x^*}{K+b}\right\}\mbox{\boldmath $1$}$, where $b\ge 1$ is a pre-determined design parameter, and $p_{\max}$ is given by
\begin{equation}
 p_{\max} = \min \left\{1, \frac{x^*}{J_{\epsilon_v}+b}\right\}.
\end{equation}
We intend to design a distributed MAC algorithm to maximize $U(K, p, \{C_{rj} \})$ by maintaining channel contention at a desired level. Recall that channel contention level is measured by $q_v$, termed the ``channel contention measure'', which is the success probability of the virtual packet. $q_v(\mbox{\boldmath $p$},K)$ is a function of the transmission probability vector $\mbox{\boldmath $p$}$ and user number $K$. Because $q_v(\mbox{\boldmath $p$},K)$ equals the summation of a finite number of polynomial terms, it should be Lipschitz continuous in $\mbox{\boldmath $p$}$ for any finite $K$. When all users transmit with the same probability $p$, i.e., $\mbox{\boldmath $p$}=p \mbox{\boldmath $1$}$, we also write $q_v$ as
\begin{equation}
 q_v(p, K) = \sum_{j=0}^{K} \left( {K \atop j} \right)  p^{j} (1-p)^{K-j} C_{vj}.
 \label{q_vSingle}
\end{equation}
We assume that $q_v$ should be measured at the receiver, and be fed back to all users. Upon receiving $q_v$, each user should first obtain a user number estimate $\hat{K}$, and then set the corresponding transmission probability target at $\tilde{p}=\hat{p}=\min \left\{p_{\max}, \frac{x^*}{\hat{K} +b}\right\}$, where $x^*$ is obtained by (\ref{xstar}).

Convergence of the distributed MAC algorithm, which will be presented later, depends on two key monotonicity properties. The first one is presented in the following theorem.
\begin{theorem} \label{MonotonicityOfqv}
 	 	With $C_{vj} \geq C_{v(j+1)}$, for all $j\ge 0$,  $q_v(p,K)$ defined in (\ref{q_vSingle}) is non-increasing in $p$, i.e. $\frac{\partial q_v(p,K)}{\partial p} \leq 0$. Furthermore $\frac{\partial q_v(p,K)}{\partial p} < 0$ holds with strict inequality for $K > J_{\epsilon_v}$ and $p \in (0,1)$.
\end{theorem}

The proof of Theorem \ref{MonotonicityOfqv} is given in Appendix \ref{MonotonictyOfqvProof}.

Next, we introduce the ``theoretical channel contention measure'', denoted by $q_v^*$, to characterize the desired channel contention level of the system. Let $\hat{K}$ be the user number estimate, $\hat{p} = \frac{x^*}{\hat{K}+b}$, and $N=\lfloor \hat{K}\rfloor$ be the largest integer below $\hat{K}$. We define a continuous function $q_v^*(\hat{p})$, which can also be viewed as a function of $\hat{K}$, as follows.
\begin{equation}
 q_v^*(\hat{p}) = \frac{\hat{p} - p_{N+1}}{p_N - p_{N+1}} q_N(\hat{p}) + \frac{ p_{N} - \hat{p}}{p_N - p_{N+1}} q_{N+1}(\hat{p}),
 \label{q_vstarSingle}
\end{equation}
where $p_N = \min \{p_{\max}, \frac{x^*}{N+b}\}$, $p_{N+1} = \min \{p_{\max}, \frac{x^*}{N+1+b}\}$,
\begin{equation}
 q_N(p) = \sum_{j=0}^{N}\left( {N \atop j}\right) p^j (1-p)^{N-j} C_{vj},
\end{equation}
and
\begin{equation}
 q_{N+1}(p) = \sum_{j=0}^{N+1}\left( {N+1 \atop j}\right) p^j (1-p)^{N+1-j} C_{vj}.
\end{equation}
Note that, if $\hat{K}=K$ equals the actual user number, then $q_v^*(\hat{p})$ equals the actual channel contention level at the equilibrium when all users have the same transmission probability $\hat{p}$.

The following theorem gives the second key monotonicity property, which shows that, given an arbitrary $x^*>0$ and with an appropriate choice of $b$, $q_v^*(\hat{p})$ is non-decreasing in $\hat{p}$.
\begin{theorem} \label{Monotonicity_q_vstar}
If $x^* > 0$ and $b \geq \max \{1, x^*-\gamma_{\epsilon_v}\}$ with $\gamma_{\epsilon_v}$ being defined as
\begin{eqnarray}
 && \gamma_{\epsilon_v} = \min_{N, N\ge J_{\epsilon_v}, N \ge x^*-b} \nonumber \\
 && \quad \frac{\sum_{j=0}^{N} j \left( {N \atop j}\right) (\frac{p_{N+1}}{1-p_{N+1}})^j (C_{vj} - C_{v (j+1)}) }{\sum_{j=0}^{N}  \left( {N \atop j}\right) (\frac{p_{N+1}}{1-p_{N+1}})^j (C_{vj} - C_{v (j+1)})},
  	\label{Gamma}
 \end{eqnarray}
then $q_v^*(\hat{p})$ defined in (\ref{q_vstarSingle}) is non-decreasing in $\hat{p}$. Furthermore, if $b > \max \{1, x^*-\gamma_{\epsilon_v}\}$ holds with strict inequality, then $q_v^*(\hat{p})$ is strictly increasing in $\hat{p}$ for $\hat{p} \in (0,p_{\max})$.
\end{theorem}

The proof of Theorem \ref{Monotonicity_q_vstar} is given in Appendix \ref{Monotonicity_q_vstarProof}. Note that, if $\epsilon_v$ is small enough to satisfy $C_{vj}=C_{v(j+1)}$ for all $j< J_{\epsilon_v}$, then we have $\gamma_{\epsilon_v}=J_{\epsilon_v}$. Otherwise, $\gamma_{\epsilon_v} \le J_{\epsilon_v}$ is generally true.

We are now ready to present the distributed MAC algorithm.

\textbf{Distributed MAC algorithm}
\begin{enumerate}
	\item Each user initializes its transmission probability.
 	\item \label{SingleOptionStep2} Over an interval of $Q$ time slots, with $Q\ge 1$, the receiver measures the success probability of the virtual packet, denoted by $q_v$, and feeds $q_v$ back to all transmitters.
 	\item Upon receiving $q_v$, each user derives a probability target $\hat{p}$ by solving the following equation.
 	\begin{equation}
 	q_v^*(\hat{p}) = q_v.
 	\label{Calculating_phat}
 	\end{equation}
	If a $\hat{p} \in [0,p_{\max}]$ satisfying (\ref{Calculating_phat}) cannot be found, each user sets $\hat{p}$ at $\hat{p} = p_{\max}$ if $q_v > q_v^*(p_{\max})$, or at $\hat{p} = 0$ if $q_v < q_v^*(0)$.
 	\item Each user, say user $k$, then updates its transmission probability $p_k$ by
 	\begin{equation}
 	p_k = (1-\alpha) p_k + \alpha \hat{p}.
 	\label{UpdatingProbability}
	\end{equation}
 	\item The process is repeated from Step \ref{SingleOptionStep2} till probabilities of all users converge.
\end{enumerate}

Convergence of the proposed MAC algorithm is stated in the following theorem.
\begin{theorem} \label{ConvergenceMACAlgorithm}
    Consider the $K$-user multiple access network presented in this section. Let $x^* > 0$ and $\epsilon_v >0$. Let $b$ be chosen to satisfy $b > \max \{1, x^*-\gamma_{\epsilon_v}\}$, where $\gamma_{\epsilon_v}$ is defined in (\ref{Gamma}). With the proposed MAC algorithm, the associated ODE given in (\ref{AssociatedSystemODE}) has a unique equilibrium at $\mbox{\boldmath $p$}^* = \min \left\{p_{\max}, \frac{x^*}{K+b}\right\}\mbox{\boldmath $1$}$. Furthermore, probability target $\hat{p}(\mbox{\boldmath $p$})$ as a function of the transmission probability vector $\mbox{\boldmath $p$}$ satisfies Conditions \ref{MeanBiasAssumption} and \ref{LipschitzAssumption}. Consequently, transmission probability vector $\mbox{\boldmath $p$}$ converges to $\mbox{\boldmath $p$}^*$ in the sense specified in Theorems \ref{StochasticProbabilityOneConvergence} and \ref{StochasticWeakConvergence}.
\end{theorem}

The proof of Theorem \ref{ConvergenceMACAlgorithm} is given in Appendix \ref{ProofConvergenceMACAlgorithm}.

Theorem \ref{ConvergenceMACAlgorithm} indicates that, so long as $b$ satisfies $b > \max \{1, x^*-\gamma_{\epsilon_v}\}$ where $\gamma_{\epsilon_v}=J_{\epsilon_v}$ if $\epsilon_v$ is small enough, convergence of the MAC algorithm is guaranteed despite of the coding parameters of the virtual packet. However, one should note that optimality of the MAC algorithm depends on $b$ and $J_{\epsilon_v}$, both of which depend on the virtual channel parameter set $\{C_{vj}\}$ and hence the design choice of the virtual packet. Assume that setting the transmission probabilities of all users at $p=\min\left\{1, \frac{x^*}{K}\right\}$ should be an ideal choice for optimizing the chosen utility\footnote{Note that this is indeed optimal for sum throughput maximization over a collision channel \cite{ref Hajek85}\cite{ref Ghez88}.}. Because the proposed MAC algorithm sets system equilibrium at $\mbox{\boldmath $p$}^*=\min\left\{p_{\max}, \frac{x^*}{K+b}\right\}\mbox{\boldmath $1$}$, there are two optimality concerns. On one hand, when user number $K$ takes a large value, it is a general preference that one should design the virtual packet to allow a relatively small value of $b$, which implies that $\gamma_{\epsilon_v}$ and $J_{\epsilon_v}$ should not be much smaller than $x^*$. On the other hand, when user number $K$ takes a small value, one would want to have $p_{\max}$ get close to $1$. This means that $J_{\epsilon_v}$ also should not be much larger than $x^*$. Taking both optimality concerns into consideration, a general guideline is to design coding parameters of the virtual packet such that $J_{\epsilon_v}$ (and $\gamma_{\epsilon_v}$) should be slightly smaller than $x^*$, and $b$ should be close to $1$.

\section{Multiple Transmission Options}
\label{MultipleOptions}
In this section, we consider the general situation when each user is equipped with $M\ge 1$ transmission options plus an idling option. Each user $k$, $k=1, \dots, K$, should maintain an $M$-length transmission probability vector $\mbox{\boldmath $p$}_k=p_k \mbox{\boldmath $d$}_k$, where $p_k$ is the transmission probability and $\mbox{\boldmath $d$}_k$ is the transmission direction vector. We use an $MK$-length vector $\mbox{\boldmath $P$}= [\mbox{\boldmath $p$}_1^T, \dots, \mbox{\boldmath $p$}_K^T]^T$ to list the transmission probability vectors of all users. We still assume that there is one virtual packet being transmitted in each time slot, and virtual packets transmitted in different time slots are identical. With a general channel model, the objective is again to develop a distributed MAC algorithm to maximize a chosen symmetric network utility.

We use two sets of channel parameter functions to model the link layer channel. Assume that all users should have the same transmission direction vector $\mbox{\boldmath $d$}$. We define $\{C_{rij} (\mbox{\boldmath $d$}) \} $ for $1 \leq i \leq M$ and $j \geq 0$ as the ``real channel parameter function set". $C_{rij}(\mbox{\boldmath $d$})$ is the conditional success probability of a real packet corresponding to the $i$th transmission option, should the packet be transmitted in parallel with $j$ other real packets. We also define $\{C_{vj} (\mbox{\boldmath $d$}) \}$ for $j \geq 0$ as the ``virtual channel parameter function set". $C_{vj}(\mbox{\boldmath $d$})$ is the success probability of the virtual packet should it be transmitted in parallel with $j$ real packets. We assume that $C_{vj}(\mbox{\boldmath $d$})\ge C_{v(j+1)}(\mbox{\boldmath $d$})$ should hold for all $j\ge 0$ and for any $\mbox{\boldmath $d$}$. That is, with users having the same transmission direction vector $\mbox{\boldmath $d$}$, if the number of parallel real packet transmissions increases, the chance of a virtual packet getting through the channel should not increase. Let $\epsilon_v> 0$ be a pre-determined constant. We define $J_{\epsilon_v} (\mbox{\boldmath $d$})$ as the smallest integer such that $C_{v J_{\epsilon_v}}(\mbox{\boldmath $d$})$ is strictly larger than $C_{v (J_{\epsilon_v}+1)}(\mbox{\boldmath $d$})+ \epsilon_v$, i.e.,
\begin{equation}
J_{\epsilon_v} (\mbox{\boldmath $d$}) =\mathop{\arg \min}_j C_{vj} (\mbox{\boldmath $d$}) > C_{v(j+1)}(\mbox{\boldmath $d$}) +\epsilon_v.
\label{DefiningJ_epsilon_v}
\end{equation}
By definition, $J_{\epsilon_v} (\mbox{\boldmath $d$})$ is a function of $\mbox{\boldmath $d$}$. Because both $\{C_{rij} (\mbox{\boldmath $d$}) \} $ and $\{C_{vj} (\mbox{\boldmath $d$}) \}$ can be derived from the physical layer channel model and the coding parameters of the packets \cite{ref Luo12}\cite{ref Wang12}\cite{ref Luo15}, we assume that they should be known at the transmitters and at the receiver. Note that, while $\{C_{vj} (\mbox{\boldmath $d$}) \}$ depends on the coding detail of the virtual packet, virtual packet is not involved in the calculation of $\{C_{rij} (\mbox{\boldmath $d$}) \} $.

We assume that users intend to maximize a symmetric utility function. Under the assumption that all users should have the same transmission probability vector (at the equilibrium), the utility function $U(K, \mbox{\boldmath $p$}, \{C_{rij}(\mbox{\boldmath $d$}) \})$ is defined as a function of the user number $K$, the common transmission probability vector $\mbox{\boldmath $p$} = p \mbox{\boldmath $d$}$, and the real channel parameter function set $\{C_{rij}(\mbox{\boldmath $d$}) \}$. For example, suppose that users intend to maximize the symmetric sum throughput of the network. If the $i$th transmission option has a communication rate of $r_i$ (bits/time slot), then the utility function should be given by
\begin{eqnarray}
&& U(K, \mbox{\boldmath $p$}, \{C_{rij}(\mbox{\boldmath $d$}) \}) = K \sum_{i=1}^{M} d_i r_i\sum_{j=0}^{K-1} \left( {K-1 \atop j} \right) \nonumber \\
&& \quad \times p^{j+1} (1-p)^{K-1-j} C_{rij}(\mbox{\boldmath $d$}).
\label{Throughput}
\end{eqnarray}

We intend to design a distributed MAC algorithm to maximize the chosen utility $U(K, \mbox{\boldmath $p$}, \{C_{rij}(\mbox{\boldmath $d$}) \})$ by maintaining channel contention measure $q_v$ at a desired level. $q_v(\mbox{\boldmath $P$}, K)$ is a function of user number $K$ and the $MK$-length transmission probability vector $\mbox{\boldmath $P$}$. Because $q_v(\mbox{\boldmath $P$}, K)$ equals the summation of a finite number of polynomial terms, it is Lipschitz continuous in $\mbox{\boldmath $P$}$ for any finite $K$. When all users have the same transmission probability vector $\mbox{\boldmath $p$}=p\mbox{\boldmath $d$}$, we have $\mbox{\boldmath $P$}=\mbox{\boldmath $1$}\otimes\mbox{\boldmath $p$}$. In this case, we also write $q_v$ as
\begin{equation}
q_v(\mbox{\boldmath $p$}, K)=\sum_{j=0}^K {K \choose j}p^j(1-p)^{K-j}C_{vj}(\mbox{\boldmath $d$}).
\end{equation}
Upon obtaining $q_v$ from the receiver, we assume that each user should first derive a user number estimate $\hat{K}$ by comparing $q_v$ with a ``theoretical channel contention measure'' $q_v^*(\hat{K})$, which is a function of $\hat{K}$. A user should then set its target transmission probability vector $\mbox{\boldmath $\hat{p}$}$ according to a designed vector parameter function $\mbox{\boldmath $p$}(\hat{K})$. Next, we use an example to illustrate the desired properties of the parameter function $\mbox{\boldmath $p$}(\hat{K})$.

\begin{example} \label{Example1}
		Consider a time-slotted multiple access network over a multi-packet reception channel. Each user is equipped with two transmission options respectively labeled as the high-rate option and the low-rate option. If all packets are encoded using the low-rate option, then the channel can support the parallel transmissions of no more than $12$ packets. We assume that one packet from the high-rate option is equivalent to the combination of $4$ low-rate packets. That is, the channel can support the parallel transmissions of $n_h$ high-rate packets plus $n_l$ low-rate packets if and only if $\frac{1}{3}n_h+\frac{1}{12}n_l \leq 1$. The utility function is chosen to be the sum system throughput. Suppose that all users should hold the same transmission probability vector $\mbox{\boldmath $p$} = [p_h, p_l]^T$ where $p_h$ and $p_l$ denote the probabilities of a user choosing the high-rate option and the low-rate option, respectively. We obtain the optimum probability vector as $\mbox{\boldmath $p$}^*=[p^*_h, p^*_l]^T = \mathop{\arg\max}_{\mbox{\scriptsize \boldmath $p$}} U(K, \mbox{\boldmath $p$}, \{C_{rij} (\mbox{\boldmath $d$})\})$. Figure \ref{OptimalTransmissionProbability} illustrates $p^*_h$ and $p^*_l$ as functions of the user number.
    \begin{figure}[ht]
        \begin{center}
            \includegraphics[width=3 in]{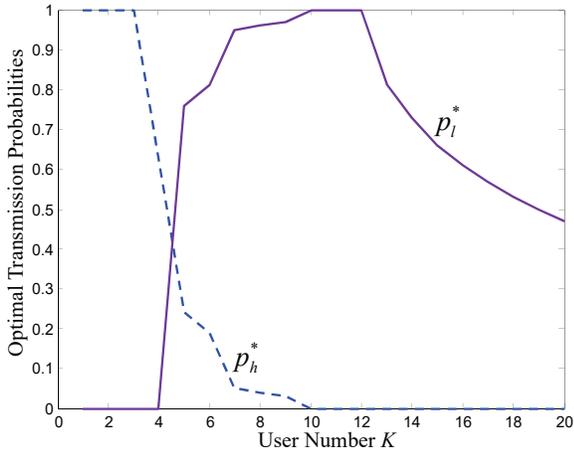}\caption{\label{OptimalTransmissionProbability}  Optimal transmission probabilities of a $K$-user multiple access system with each user having two transmission options.}
        \par\end{center}
    \end{figure}
We can see that, if we write $\mbox{\boldmath $p$}^*=p^* \mbox{\boldmath $d$}^*$, then $\mbox{\boldmath $d$}^*$ is fixed at $\mbox{\boldmath $d$}^*=[1, 0]^T$ for $K\le 4$, and fixed at $\mbox{\boldmath $d$}^*=[0, 1]^T$ for $K\ge 10$. $\mbox{\boldmath $d$}^*$ transits from $[1, 0]^T$ to $[0, 1]^T$ in the region of $4\le K\le 10$.
\end{example}

According to the above observation, we assume that the vector parameter function $\mbox{\boldmath $p$}(\hat{K})$ should be designed to satisfy the following properties termed the ``Head and Tail Condition''.
\begin{condition}\label{HeadTailAssumption}(Head and Tail)
		Let $\epsilon_v > 0$ be a pre-determined constant. Let $J_{\epsilon_v}$ be defined in (\ref{DefiningJ_epsilon_v}). There exist two integer-valued constants $0 < \underline{K} \leq \overline{K}$, such that,
		\begin{enumerate}
			\item $\underline{K} \ge J_{\epsilon_v}(\mbox{\boldmath $d$}(\underline{K}))$ and $\mbox{\boldmath $d$}(\hat{K})= \mbox{\boldmath $d$}(\underline{K})$ for $\hat{K} \leq \underline{K}$.
			\item $\overline{K} > J_{\epsilon_v}(\mbox{\boldmath $d$}(\overline{K}))$ and $\mbox{\boldmath $d$}(\hat{K})= \mbox{\boldmath $d$}(\overline{K})$ for $\hat{K} \geq \overline{K}$.
		\end{enumerate}
\end{condition}

The Head and Tail Condition indicates that, when $\hat{K}$ is either small enough or large enough, $\mbox{\boldmath $d$}(\hat{K})$ should stop changing in $\hat{K}$. Note that, when $\mbox{\boldmath $d$}(\hat{K})$ is fixed at a constant vector, the system becomes equivalent to one with each user having a single transmission option. The virtual channel parameter set of the equivalent system is given by $\{C_{vj}\}=\{C_{vj}(\mbox{\boldmath $d$})\}$. The real channel parameter set of the equivalent system, on the other hand, depends on the chosen utility function. For example, if the utility function is the sum throughput given in (\ref{Throughput}), then the equivalent real channel parameter set $\{C_{rj}\}$ should be obtained by $C_{rj}=\sum_{i=1}^M d_ir_iC_{rij}(\mbox{\boldmath $d$})$, for $j\ge 0$. We assume that, in the head regime when $\hat{K}\le \underline{K}$ and the tail regime when $\hat{K}\ge \overline{K}$, core parameter functions of the MAC algorithm, including the theoretical channel contention measure $q_v^*(\hat{K})$ and the probability target function $p(\hat{K})$, should be designed according to the guideline presented in Section \ref{SingleOption}. The detail is not repeated here.

Let us temporarily assume that the vector parameter function $\mbox{\boldmath $p$}(\hat{K})$ has been completely determined, not only for $\hat{K}\le \underline{K}$ and $\hat{K}\ge \overline{K}$, but also for $\underline{K}\le \hat{K} \le \overline{K}$. To present the distributed MAC algorithm, we need to define the theoretical channel contention measure $q_v^*(\hat{K})$ as follows. Let $N= \lfloor \hat{K} \rfloor$ be the largest integer below $\hat{K}$. For $\hat{K} \leq \underline{K}$ and $\hat{K} \geq \overline{K}$, we define $q_v^*(\hat{K})$ as,
\begin{eqnarray}
&& q_v^*( \hat{K} ) = \frac{p(\hat{K})-p(N+1)}{p(N)-p(N+1)} q_v( \mbox{\boldmath $p$}(\hat{K}),N ) \nonumber \\
&& \quad + \frac{p(N)-p(\hat{K})}{p(N)-p(N+1)} q_v( \mbox{\boldmath $p$}(\hat{K}),N+1 ),
\label{qvstarFixedD}
\end{eqnarray}
which is consistent with (\ref{q_vstarSingle}). For $\underline{K} < \hat{K} < \overline{K}$, we define $q_v^*(\hat{K})$ as,
\begin{eqnarray}
&& q_v^*( \hat{K} ) =(N+1-\hat{K}) q_v( \mbox{\boldmath $p$}(\hat{K}),N ) \nonumber \\
&& \quad +(\hat{K}-N) q_v( \mbox{\boldmath $p$}(\hat{K}),N+1 ).
\label{qvstarVariableD}
\end{eqnarray}
According to (\ref{qvstarFixedD}) and (\ref{qvstarVariableD}), $q_v^*(\hat{K})$ equals $q_v(\mbox{\boldmath $p$}(\hat{K}), \hat{K})$ for integer-valued $\hat{K}$. For non-integer-valued $\hat{K}$, $q_v^*(\hat{K})$ is a linear interpolation between $q_v( \mbox{\boldmath $p$}(\hat{K}),N )$ and $q_v( \mbox{\boldmath $p$}(\hat{K}), N+1 )$. Note that the interpolation approach used for $\hat{K} \leq \underline{K}$ and $\hat{K} \geq \overline{K}$ is different from the one used for $\underline{K} < \hat{K} < \overline{K}$.

Next, we present the distributed MAC algorithm below.

\textbf{Distributed MAC Algorithm:}
\begin{enumerate}
	\item Each user initializes its transmission probability vector.
	\item \label{Step2} Let $Q>0$ be a pre-determined integer. Over an interval of $Q$ time slots, the receiver measures the success probability of the virtual packet, denoted by $q_v$, and feeds $q_v$ back to all users.
	\item Upon receiving $q_v$, each user derives a user number estimate $\hat{K}$ by solving the following equation.
	\begin{equation}
	q_v^*(\hat{K}) = q_v.
	\label{EstimatingUserNumber}
	\end{equation}
	If a $\hat{K}$ satisfying (\ref{EstimatingUserNumber}) cannot be found, users set $\hat{K} = J_{\epsilon_v}(\mbox{\boldmath $d$}(\underline{K}))$ if $q_v > q_v^*(J_{\epsilon_v}(\mbox{\boldmath $d$}(\underline{K})))$, or set $\hat{K} = \infty$ otherwise.
	\item Each user, say user $k$, updates its transmission probability vector by
	\begin{equation}
	\mbox{\boldmath $p$}_k= (1-\alpha) \mbox{\boldmath $p$}_k + \alpha \mbox{\boldmath $p$}(\hat{K}),
	\label{UpdateProbabilityVector}
	\end{equation}
	where $\alpha$ is the step size parameter for user $k$.
	\item The process is repeated from Step \ref{Step2} till transmission probability vectors of all users converge.
\end{enumerate}

We intend to design the distributed MAC algorithm with the following convergence property. If $K\ge J_{\epsilon_v}(\mbox{\boldmath $d$}(\underline{K}) )$, we intend to have $\hat{K}=K$ at the equilibrium. While if $K<J_{\epsilon_v}(\mbox{\boldmath $d$}(\underline{K}) )$, we intend to have $\hat{K}=J_{\epsilon_v}(\mbox{\boldmath $d$}(\underline{K}) )$ at the equilibrium. To ensure convergence of the MAC algorithm, we require that the vector parameter function $\mbox{\boldmath $p$}(\hat{K})$ and the theoretical channel contention measure $q_v^*(\hat{K})$ should satisfy the following ``Monotonicity and Gradient Condition'' for $\underline{K}\le \hat{K} \le \overline{K}$.

\begin{condition} \label{MonotoncityGradientCondition} (Monotonicity and Gradient)
	For $\underline{K} \le \hat{K} \le \overline{K}$,
	\begin{enumerate}
		\item \label{GradientConditionItem1} $\mbox{\boldmath $p$}(\hat{K}) = p(\hat{K}) \mbox{\boldmath $d$}(\hat{K})$ should be Lipschitz continuous in $\hat{K}$, i.e., there exists a constant $K_g > 0$, such that the following inequality is satisfied for all $\hat{K}_a, \hat{K}_b \in [\underline{K}, \overline{K}]$.
		\begin{equation}
		\| \mbox{\boldmath $p$}(\hat{K}_a) - \mbox{\boldmath $p$}(\hat{K}_b) \| \leq K_g |\hat{K}_a - \hat{K}_b|.
		\label{Lipschitzp}
		\end{equation}
		\item \label{GradientConditionItem2} $q_v^*(\hat{K})$ should be continuous and strictly decreasing in $\hat{K}$. There exists a positive constant $\epsilon_q$, such that for all $\hat{K}_a, \hat{K}_b \in [\underline{K}, \overline{K}]$, we have
		\begin{equation}
		|q_v^*(\hat{K}_a)-q_v^*(\hat{K}_b)| \geq \epsilon_q |\hat{K}_a - \hat{K}_b|.
		\label{Lipschitzqvstar}
		\end{equation}
		\item \label{GradientConditionItem3} There exists a constant $\epsilon_v>0$, such that $\hat{K} > J_{\epsilon_v}(\mbox{\boldmath $d$}(\hat{K}))$ should be satisfied for all $\hat{K} \in [\underline{K}, \overline{K}]$.
		\item \label{GradientConditionItem4} There exist $0 < \underline{p} < \overline{p} <1$ to satisfy $\underline{p} \leq p(\hat{K}) \leq \overline{p}$ for all $\hat{K} \in [\underline{K}, \overline{K}]$.
	\end{enumerate}
\end{condition}

Convergence property of the proposed MAC algorithm under Conditions \ref{HeadTailAssumption} and \ref{MonotoncityGradientCondition} is stated in the following theorem.
\begin{theorem} \label{TheoremOfConvergence2}
	Consider a multiple access system with $K$ users adopting the proposed distributed MAC algorithm to update their transmission probability vectors. Suppose that Condition \ref{HeadTailAssumption} is satisfied. Let the vector parameter function $\mbox{\boldmath $p$}(\hat{K})$ and the theoretical channel contention measure $q_v^*(\hat{K})$ be designed according to the guideline described in Section \ref{SingleOption} for $\hat{K} \le \underline{K}$ and $\hat{K} \ge \overline{K}$. Assume that $\mbox{\boldmath $p$}(\hat{K})$ and $q_v^*(\hat{K})$ are designed to satisfy Condition \ref{MonotoncityGradientCondition} for $\underline{K} \le \hat{K} \le \overline{K}$. Then, the associated ODE defined in (\ref{AssociatedSystemODE}) has a unique equilibrium at $\mbox{\boldmath $P$}^*= \mbox{\boldmath $1$} \otimes \mbox{\boldmath $p$}(K)$. The probability target $\hat{\mbox{\boldmath $p$}}(\mbox{\boldmath $P$})$ as a function of the transmission probability vector $\mbox{\boldmath $P$}$ satisfies Conditions \ref{MeanBiasAssumption} and \ref{LipschitzAssumption}. Consequently, transmission probability vectors of all users should converge to $\mbox{\boldmath $P$}^*= \mbox{\boldmath $1$} \otimes \mbox{\boldmath $p$}(K)$ in the sense specified in Theorems \ref{StochasticProbabilityOneConvergence} and \ref{StochasticWeakConvergence}.
\end{theorem}

The proof of Theorem \ref{TheoremOfConvergence2} is given in Appendix \ref{ProofofTheoremOfConvergence2}.

Compared with the monotonicity conditions presented in Section \ref{SingleOption}, while Condition \ref{MonotoncityGradientCondition} still requires $q_v^*(\hat{K})$ to be strictly decreasing in $\hat{K}$,  $q_v(\mbox{\boldmath $p$}(\hat{K}), K)$ is no longer required to be increasing in $\hat{K}$ for a given $K$. Furthermore, unlike the single transmission option case where $p(\hat{K})$ as a function of $\hat{K}$ is completely specified in a closed form, Condition \ref{MonotoncityGradientCondition} did not explain how $\mbox{\boldmath $p$}(\hat{K})$ should be designed to possess the required properties.

Next, we show that, so long as one can manually design $\mbox{\boldmath $p$}(\hat{K})$ for a set of chosen points with integer-valued $\hat{K}$ to satisfy the following ``Pinpoints Condition'', then there is a simple approach to complete the $\mbox{\boldmath $p$}(\hat{K})$ function to satisfy Condition \ref{MonotoncityGradientCondition} for $\underline{K} \le \hat{K} \le \overline{K}$.

\begin{condition}\label{PinPointsAssumption} (Pinpoints)
	Let $\hat{K}_i$ for $i=0, \dots, L$ be $L+1$ integers such that $\underline{K}= \hat{K}_0 < \hat{K}_1 < \dots < \hat{K}_L=\overline{K}$. For $i=0,\dots,L$ and $0 \leq \lambda <1$, define
	\begin{eqnarray}
	&& \hat{K}_{i\lambda} = (1-\lambda) \hat{K}_{i-1} + \lambda \hat{K}_i, \nonumber \\
	&& \mbox{\boldmath $d$}_{i \lambda} = (1-\lambda) \mbox{\boldmath $d$}(\hat{K}_{i-1}) + \lambda \mbox{\boldmath $d$}(\hat{K}_i), \nonumber \\
	&& q^*_{vi\lambda} = (1-\lambda) q^*_v(\hat{K}_{i-1})  + \lambda q^*_v(\hat{K}_i).
	\label{DefinitionOfNonPinnedPoints}
	\end{eqnarray}
	We have the following conditions.
	\begin{enumerate}
		\item \label{Item1Condition5} There exists a positive constant $\epsilon_q$ to satisfy $q_v^*(\hat{K}_{i-1}) - q_v^*(\hat{K}_{i}) \geq \epsilon_q$, for all $i=1, \dots, L$.
		\item There exists a constant $\epsilon_v>0$, such that for all $i=1, \dots, L$ and for all $0\le \lambda <1$, we have $\hat{K}_{i\lambda} > J_{\epsilon_v}(\mbox{\boldmath $d$}_{i \lambda})$, where $J_{\epsilon_v}(\mbox{\boldmath $d$}_{i \lambda})$ is defined in (\ref{DefiningJ_epsilon_v}).
		\item There exist $0 < \underline{p} < \overline{p} <1$ to satisfy $\underline{p} \leq p(\hat{K}_i) \leq \overline{p}$ for all $i=1, \dots, L$.
		\item \label{Item4Condition5} Extend the definition of $q_v(\mbox{\boldmath $p$}, \hat{K})$ for non-integer-valued $\hat{K}$ as
		\begin{eqnarray}
		&& q_v(\mbox{\boldmath $p$}, \hat{K}) = (\lfloor \hat{K} \rfloor +1-\hat{K}) q_v(\mbox{\boldmath $p$}, \lfloor \hat{K} \rfloor) \nonumber \\
		&& \quad + (\hat{K}- \lfloor \hat{K} \rfloor) q_v(\mbox{\boldmath $p$}, \lfloor \hat{K} \rfloor+1),
		\label{DefinitionOfq_vNoninteger}
		\end{eqnarray}
        The following inequality should be satisfied for all $i=1, \dots, L$ and for all $0\le \lambda <1$.
        \begin{equation}
        q_v(\overline{p} \mbox{\boldmath $d$}_{i \lambda}, \hat{K}_{i \lambda}) \leq q^*_{vi \lambda} \leq q_v(\underline{p} \mbox{\boldmath $d$}_{i \lambda}, \hat{K}_{i \lambda}).
        \end{equation}
	\end{enumerate}
\end{condition}

With $\mbox{\boldmath $p$}(\hat{K})$ being designed for the $L+1$ pinpoints, we propose the following interpolation approach to complete $\mbox{\boldmath $p$}(\hat{K})$ for $\underline{K} \leq \hat{K} \leq \overline{K}$.

\textbf{Interpolation Approach}
Assume that $\mbox{\boldmath $p$}(\hat{K})$ is designed for $\hat{K}_i$, $i=0, .... L$, with $\underline{K}= \hat{K}_0 < \hat{K}_1 < \dots < \hat{K}_L=\overline{K}$, to satisfy Condition \ref{PinPointsAssumption}. For $i=1, \dots, L$ and for all $0\le \lambda <1$, let $\hat{K}_{i\lambda}$, $\mbox{\boldmath $d$}_{i \lambda}$ and $q^*_{vi\lambda}$ be defined in (\ref{DefinitionOfNonPinnedPoints}). Let $q_v(\mbox{\boldmath $p$}, \hat{K})$ be defined in (\ref{DefinitionOfq_vNoninteger}). We choose $p(\hat{K}_{i \lambda})$ to satisfy the following equality.
\begin{equation}
q_v(p(\hat{K}_{i \lambda}) \mbox{\boldmath $d$}_{i \lambda}, \hat{K}_{i\lambda}) = q^*_{vi\lambda}.
\label{FindingpK}
\end{equation}
This leads to $\mbox{\boldmath $p$}(\hat{K}_{i \lambda})= p(\hat{K}_{i \lambda}) \mbox{\boldmath $d$}_{i \lambda}$. Note that the existence of a solution to (\ref{FindingpK}) is guaranteed by Item \ref{Item4Condition5} of Condition \ref{PinPointsAssumption}.

Effectiveness of the Interpolation Approach is stated in the following theorem.
\begin{theorem} \label{PinPointsTheorem}
    Assume that $\mbox{\boldmath $p$}(\hat{K})$ is designed for a set of $L+1$ pinpoints $\{\hat{K}_i\}$, $i=0, \dots, L$, with $\underline{K}=\hat{K}_0 < \hat{K}_1, \dots, < \hat{K}_L=\overline{K}$, to satisfy Condition \ref{PinPointsAssumption}. After completing the function using the Interpolation Approach, $\mbox{\boldmath $p$}(\hat{K})$ and $q_v^*(\hat{K})$ functions satisfy the Monotonicity and Gradient Condition \ref{MonotoncityGradientCondition} for $\underline{K} \le \hat{K} \le \overline{K}$.
\end{theorem}

The proof of Theorem \ref{PinPointsTheorem} is given in appendix \ref{ProofOfPinPointsTheorem}.

Note that, in the single transmission option case discussed in Section \ref{SingleOption}, $p(\hat{K})$ is specified in a closed form with a small number of design parameters. Monotonicity property of $q_v^*(\hat{K})$ is proven theoretically. With multiple transmission options, however, such a direct-design approach faces a key challenge. Due to generality of the system model, when $\mbox{\boldmath $d$}(\hat{K})$ changes in $\hat{K}$ and consequently affects the channel parameters, it is often difficult to theoretically characterize its impact on the $q_v^*(\hat{K})$ function. Alternatively, we switched to a search-assisted approach to first manually design $\mbox{\boldmath $p$}(\hat{K})$ for a set of pinpoints to satisfy Condition \ref{PinPointsAssumption}, and then to use the Interpolation Approach to complete the $\mbox{\boldmath $p$}(\hat{K})$ function to satisfy Condition \ref{MonotoncityGradientCondition}. Because the Interpolation Approach only ensures convergence of the proposed MAC algorithm but pays no attention to its optimality, one often needs to carefully adjust the design of the pinpoints to direct the $\mbox{\boldmath $p$}(\hat{K})$ function toward a near optimal solution.

\section{Simulation Results}
\label{Simulation}
In this section, we provide computer examples to illustrate both the optimality and the convergence properties of the proposed MAC algorithms.

\begin{example}\label{MultiOptionExample}
In \cite{ref Hajek85}, a similar stochastic approximation model was proposed for the maximization of symmetric sum throughput of a distributed multiple access network over a collision channel. Assume that there are $K$ users each having a single transmission option and a saturated message queue. If $K$ is known, the optimum solution that maximizes the symmetric sum throughput is to set the transmission probabilities of all users at $p_{\mbox{\scriptsize opt}}=\frac{1}{K}$ \cite{ref Hajek85}. In \cite{ref Hajek85}, under the assumption of an unknown user number and due to the constraint of certain monotonicity properties, it was suggested that equilibrium of the distributed MAC algorithm should be set at $p_a$, which is obtained by solving the following equation.
\begin{equation}
e P(\mbox{idle}) -1 - 0.5 \sqrt{p_a}=0, \quad P( \mbox{idle})=(1-p_a)^K,
\end{equation}
where $P( \mbox{idle})$ is the idling probability of the channel that can be measured without knowing the value of $K$.

Let us follow the design guideline presented in Section \ref{SingleOption} of this paper. With the collision channel model, the real channel parameter set $\{C_{rj}\}$ is given by $C_{r0}=1$ and $C_{rj}=0$ for $j>0$. With the utility chosen to be the symmetric sum throughput, we get from (\ref{xstar}) that $x^*=1$. Assume that a virtual packet should have the same coding parameters of a real packet. Consequently, the virtual channel parameter set $\{C_{vj}\}$ is given by $C_{v0}=1$ and $C_{vj}=0$ for $j>0$. Choose $\epsilon_v=0.01$, we get $\gamma_{\epsilon_v}=J_{\epsilon_v}=0$. Therefore, we can set $b=1.01 >x^*-\gamma_{\epsilon_v}$. This leads to an equilibrium with $p^*=\frac{1}{K+1.01}$.

In Figure \ref{CollisionChannelExample}, we illustrate the achieved sum throughput of the system in packet/slot as a function of the user number under the optimal transmission probability, at the equilibrium of the proposed distributed MAC algorithm, and at the equilibrium of the approach suggested in \cite{ref Hajek85}.
\begin{figure}[!h]
    \begin{center}
        \includegraphics[width=3 in]{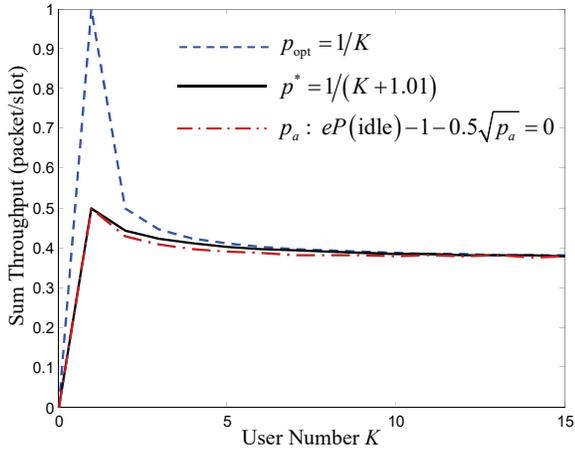}\caption{\label{CollisionChannelExample}  Sum throughput as a function of the user number for a multiple access network with a collision channel.}
    \end{center}
\end{figure}
It can be seen that, for the classical scenario presented in \cite{ref Hajek85}, the distributed MAC algorithm proposed in this paper can achieve a throughput performance better than the approach proposed in \cite{ref Hajek85}, although the improvement is indeed marginal.
\end{example}

\begin{example}
In this example, we consider distributed multiple access networking over a simple fading channel. Assume that the system has $K$ users and one receiver. Each user only has a single transmission option. In each time slot, with a probability of $0.3$, the channel can support no more than $M_1=4$ parallel real packet transmissions, and with a probability of $0.7$, the channel can support no more than $M_2=6$ parallel real packet transmissions\footnote{Such a channel can appear if there is an interfering user that transmits a packet with a probability of $0.3$ in each time slot. One packet from the interfering user is equivalent to the combination of two packets from a regular user.}. In this case, the real channel parameter set $\{C_{rj}\}$ is given by $C_{rj}=1$ for $j<4$, $C_{rj}=0.7$ for $4\le j<6$, and $C_{rj}=0$ for $j\ge 6$. Assume that users intend to maximize the symmetric system throughput weighted by a transmission energy cost of $E=0.3$. With $K$ users all transmitting at the same probability of $p$, system utility $U(K, p, \{C_{rj}\})$ is given by
\begin{eqnarray}
&& U(K, p, \{C_{rj}\})= - EKp + \nonumber \\
&& \quad \sum_{j=0}^{K-1}K{K-1 \choose j} p^{j+1}(1-p)^{K-1-j}C_{rj}.
\end{eqnarray}
Correspondingly, $x^*$ can be obtained from asymptotic utility optimization (\ref{xstar}) as $x^*=3.29$. Assume that a virtual packet should have the same coding parameters as those of a real packet. The virtual channel parameter set $\{C_{vj}\}$ is therefore identical to the real channel parameter set, i.e., $C_{vj}=C_{rj}$ for all $j\ge 0$. Choose $\epsilon_v=0.01$, we have $\gamma_{\epsilon_v}=J_{\epsilon_v}=3$. Therefore, we can set $b=1.01$.

In Figure \ref{SingleOptionExampleUtility}, we illustrate three utilities, all as functions of user number $K$.
\begin{figure}[!h]
    \begin{center}
        \includegraphics[width=3 in]{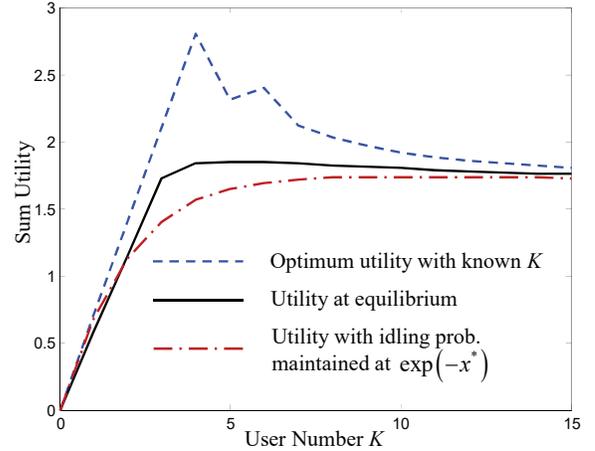}\caption{\label{SingleOptionExampleUtility} Sum utility as a function of the user number for a multiple access network over a simple fading channel.}
    \end{center}
\end{figure}
The solid curve represents the utility achieved by the proposed MAC algorithm at the designed equilibrium, with all users transmitting at a probability of $p^*=\min\left\{p_{\max}, \frac{x^*}{K+b}\right\}$. The dashed curve represents the optimum utility under the assumption that user number $K$ is known. Note that the optimum utility is not necessarily achievable without the knowledge of $K$. The dash-dotted curve represents the utility if we maintain the channel idling probability at its asymptotically optimal value of $\exp(-x^*)$, or equivalently, if we set the transmission probabilities of all users at $1-\exp\left(-\frac{x^*}{K}\right)$. This is an intuitive extension to the key idea suggested in \cite{ref Hajek85}, although a general channel model was not discussed in \cite{ref Hajek85}\footnote{Note that, other suggestions of maintaining certain variable at its asymptotically optimal value, as discussed in \cite{ref Hajek85}, do not give a better performance in this example.}.

Next, we assume that the system has $K=8$ users. Transmission probabilities of all users are initialized at $0$. In each time slot, a channel state flag is randomly generated to indicate whether the channel can support the parallel transmissions of no more than $4$ or $6$ packets. Each user also randomly determines whether a packet should be transmitted according to its own transmission probability parameter. Whether the real packets and the virtual packet can go through the channel or not is then determined using the corresponding channel model. We use the following exponential moving average approach to measure $q_v$. $q_v$ is initialized at $q_v=1$. In each time slot, $q_v$ is updated by $q_v=(1-\frac{1}{300})q_v+\frac{1}{300}I_v$, where $I_v\in \{0, 1\}$ is an indicator of the success/failure reception status of the virtual packet in the current time slot. While this is different from the approach proposed in the distributed MAC algorithm, simulations show that an exponential averaging measurement of $q_v$ can often lead the system to convergence in a relatively small number of time slots. The rest of probability updates proceeds according to the distributed MAC algorithm introduced in Section \ref{SingleOption} with a constant step size of $\alpha=0.05$. Convergence behavior of the sum utility is illustrated in Figure \ref{SingleOptionExampleConvergence}, where sum utility is also measured using the same exponential moving average approach except that initial value of the utility is set at $0$.
\begin{figure}[ht]
    \begin{center}
        \includegraphics[width=3 in]{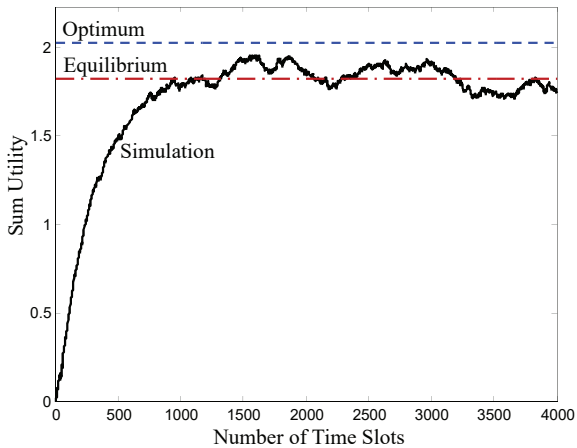}\caption{\label{SingleOptionExampleConvergence} Convergence in sum utility of a system with $K=8$ users.}
    \end{center}
\end{figure}

\end{example}

\begin{example}\label{MultipleOptionExample}
In this example, we use the system introduced in Example \ref{MultiOptionExample} to illustrate the design procedure of the $\mbox{\boldmath $p$}(\hat{K})$ function. First, we consider the ``Head'' and the ``Tail'' regimes when $\hat{K}$ is either small or large in value. We will add subscript ``H'' (or ``T'') to parameters of the ``Head'' (or the ``Tail'') regime. Without specifying the values of $\underline{K}$ and $\overline{K}$, we first determine the optimal transmission direction vectors in these two regimes as $\mbox{\boldmath $d$}_H=[1, 0]^T$ and $\mbox{\boldmath $d$}_T=[0, 1]^T$. In other words, users should only use the high rate option in the ``Head'' regime and only use the low rate option in the ``Tail'' regime. In the ``Head'' regime, the channel can support the parallel transmissions of no more than $3$ high rate packets. The real channel parameter set of the equivalent single option system is given by $\{C_{rj}\}_H$ with $C_{rj}=1$ for $j\le 2$ and $C_{rj}=0$ otherwise. By following the single option system design guideline, we get $x^*_H=\mathop{\arg\max}_x(x+x^2+\frac{x^3}{2})e^{-x}=2.27$. We design the virtual packet to be equivalent to a real high rate packet. Consequently, the virtual channel parameter set of the equivalent single option system is given by $\{C_{vj}\}_H=\{C_{rj}\}_H$. Choose $\epsilon_v=0.01$, we get $\gamma_{\epsilon_v H}=J_{\epsilon_v H}=2$, and $b_H=1.01$. In the ``Tail'' regime, on the other hand, the channel can support the parallel transmissions of no more than $12$ low rate packets. The real channel parameter set of the equivalent single option system is given by $\{C_{rj}\}_T$ with $C_{rj}=1$ for $j\le 11$ and $C_{rj}=0$ otherwise. This leads to $x^*_T=\mathop{\arg\max}_x\sum_{i=0}^{11}\frac{x^{i+1}}{i!}e^{-x}=8.82$. Because we already chose the virtual packet to be equivalent to a high rate real packet, virtual channel parameter set of the equivalent single option system in this case is given by $\{C_{vj}\}_T$ with $C_{vj}=1$ for $j\le 8$ and $C_{vj}=0$ otherwise. Therefore, with $\epsilon_v=0.01$, we have $\gamma_{\epsilon_v T}=J_{\epsilon_v T}=8$. Luckily, this supports $b_T=1.01$.

Next, we determine the values of $\underline{K}$ and $\overline{K}$. We first compare two schemes named the ``high rate option only'' scheme and the ``low rate option only'' scheme. In the ``high rate option only'' scheme, we fix $\mbox{\boldmath $d$}(\hat{K})$ at $[1, 0]^T$ for all $\hat{K}$, and set $p(\hat{K})=\min\left\{p_{\max H}, \frac{x^*_H}{\hat{K}+b_H}\right\}$, where $p_{\max H}=\frac{x^*_H}{J_{\epsilon_v H}+b_H}$. In the ``low rate option only'' scheme, we fix $\mbox{\boldmath $d$}(\hat{K})$ at $[0, 1]^T$ for all $\hat{K}$, and set $p(\hat{K})=\min\left\{p_{\max T}, \frac{x^*_T}{\hat{K}+b_T}\right\}$, where $p_{\max T}=\frac{x^*_T}{J_{\epsilon_v T}+b_T}$. By comparing the utility values and the theoretical channel contention measures of the two schemes, we choose $\underline{K}=4$ and $\overline{K}=10$.

Now consider the ``Pinpoints Condition'' for $\underline{K} \le \hat{K} \le \overline{K}$. For transmission direction vectors $\mbox{\boldmath $d$}$ satisfying $d_1>0$, with a small enough $\epsilon_v$, we generally have $J_{\epsilon_v}=2$. Therefore, so long as $\mbox{\boldmath $d$}(\hat{K})$ does not transit too quickly to $[0, 1]^T$, the condition of $\hat{K} > J_{\epsilon_v} (\mbox{\boldmath $d$}(\hat{K}))$ should hold true. Consequently, only two other key conditions need to be satisfied. The first condition is that $q_v^*(\hat{K})$ of the selected pinpoints must be strictly decreasing in $\hat{K}$. The second condition is that $p(\hat{K})$ found in the Interpolation Approach should be bounded away from $0$ and $1$. In addition, from the optimal scheme, we can see that $\mbox{\boldmath $d$}(\hat{K})$ should transit toward $[0, 1]^T$ faster than a linear transition from $\hat{K}=\underline{K}$ to $\hat{K}=\overline{K}$. With these considerations, we choose the following $4$ pinpoints. At the edge of the ``Head'' and the ``Tail'' regimes, we have $\hat{K}_0=\underline{K}=4$ with $\mbox{\boldmath $p$}(4)=\frac{x^*_H}{\underline{K}+b_H}[1, 0]^T$ and $\hat{K}_3=\overline{K}=10$ with $\mbox{\boldmath $p$}(10)=\frac{x^*_T}{\overline{K}+b_T}[0, 1]^T$. We also choose other two pinpoints at $\hat{K}_1=5$ and $\hat{K}_2=6$. We set transmission directions vectors $\mbox{\boldmath $d$}(5)$ and $\mbox{\boldmath $d$}(6)$ to be equal to the corresponding optimal transmission direction vectors, i.e., direction vectors extracted from the optimal $\mbox{\boldmath $p$}$ vectors that maximize the sum throughput at $K=5$ and $K=6$, respectively. Transmission probabilities of these two pinpoints are chosen such that the resulting $q_v^*(\hat{K})$ equals $\frac{\overline{K}-\hat{K}}{\overline{K}-\underline{K}}q_v^*(\underline{K})+\frac{\hat{K}-\underline{K}}{\overline{K}-\underline{K}}q_v^*(\overline{K})$. Note that, the purpose of designing pinpoints $\hat{K}_1=5$ and $\hat{K}_2=6$ is to help $\mbox{\boldmath $d$}(\hat{K})$ to transit quickly toward $[0, 1]^T$. The rest of the $\mbox{\boldmath $p$}(\hat{K})$ function is completed using the Interpolation Approach for $\underline{K}\le \hat{K}\le \overline{K}$. Theoretical channel contention measure $q_v^*(\hat{K})$ of the designed system is illustrated in Figure \ref{MultipleOptionExampleQvstar}.
\begin{figure}[ht]
    \begin{center}
        \includegraphics[width=3 in]{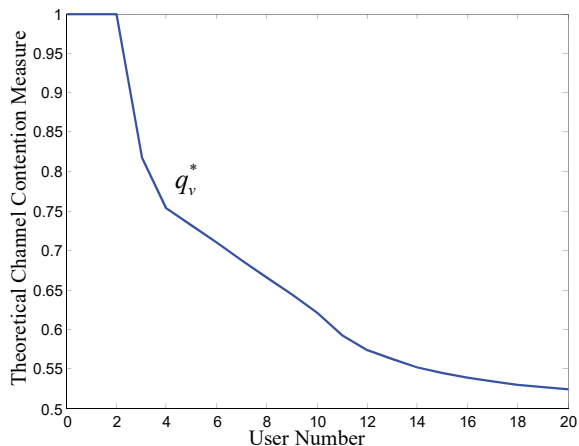}\caption{\label{MultipleOptionExampleQvstar} Theoretical channel contention measure $q_v^*$ as a function of the user number.}
    \end{center}
\end{figure}

In Figure \ref{MultipleOptionExampleThroughput}, we illustrate the theoretical sum throughput of the network as functions of user number $K$ when the transmission probability vectors of all users are set at the following four different vectors: optimal $\mbox{\boldmath $p$}(K)$ that maximizes the sum throughput, designed $\mbox{\boldmath $p$}(K)$, $\mbox{\boldmath $p$}(K)$ from the high rate option only scheme, and $\mbox{\boldmath $p$}(K)$ from the low rate option only scheme.
\begin{figure}[ht]
    \begin{center}
        \includegraphics[width=3 in]{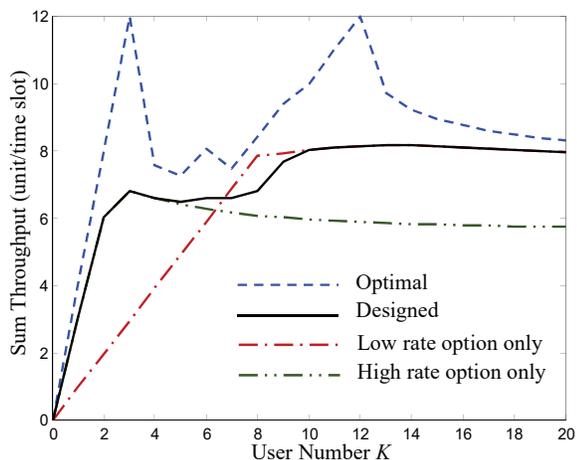}\caption{\label{MultipleOptionExampleThroughput} Sum throughput of the system as functions of the user number under different transmission probability vector settings.}
    \end{center}
\end{figure}
Assume that the two simple schemes should be reasonably good for the ``Head'' and the ``Tail'' regimes. It can be seen from Figure \ref{MultipleOptionExampleThroughput} that, with the help of the designed $\mbox{\boldmath $p$}(\hat{K})$ and $q_v^*(\hat{K})$ functions, the system can take advantage of the multiple transmission options and maintain a reasonably good performance in term of sum throughput for all user number values.

Next, we illustrate the convergence property of the distributed MAC algorithm proposed in Section \ref{MultipleOptions}. Assume that the system has $8$ users initially. Transmission probability vectors of all users are initialized at $[0,0]^T$. In each time slot, according to its own transmission probability vector, each user randomly determines whether a packet should be transmitted or not, and if the answer is positive, which transmission option should be used. The receiver measures $q_v$ using the following exponential moving average approach. $q_v$ is initialized at $q_v=1$. In each time slot, an indicator variable $I_v\in \{0, 1\}$ is used to represent the success/failure status of the virtual packet reception. $q_v$ is then updated by $q_v=(1-\frac{1}{300})q_v+\frac{1}{300}I_v$, and is fed back to the users at the end of each time slot. Each user then adapts its transmission probability vector according to the proposed MAC algorithm with a constant step size of $\alpha=0.05$.

We assume that the system experiences three stages. At Stage one, the system has $8$ users. The system enters Stage two at the $3001$th time slot, when $6$ more users enter into the system with their transmission probability vectors initialized at $[0, 0]^T$. Then at the $6001$th time slot, the system enters Stage three when $8$ users exist the system. Convergence behavior in sum throughput of the system is illustrated in Figure \ref{MultipleOptionExampleThroughputSimulation}. The corresponding optimal throughput and the theoretical throughput at the designed equilibrium are provided as references.
\begin{figure}[!ht]
    \begin{center}
        \includegraphics[width=3 in]{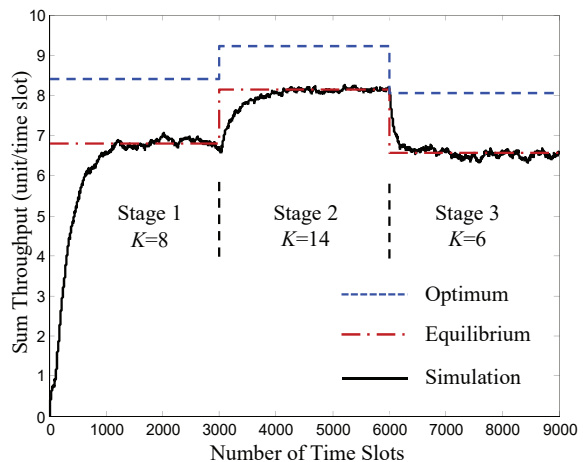}\caption{\label{MultipleOptionExampleThroughputSimulation} Convergence in sum throughput of the system. User number changed from $8$ to $14$ and then to $6$ over the three stages.}
    \end{center}
\end{figure}
In Figure \ref{MultipleOptionExampleProbabilitySimulation}, we also illustrate entries of the transmission probability vector target calculated by the users together with the corresponding theoretical values.
\begin{figure}[!ht]
    \begin{center}
        \includegraphics[width=3 in]{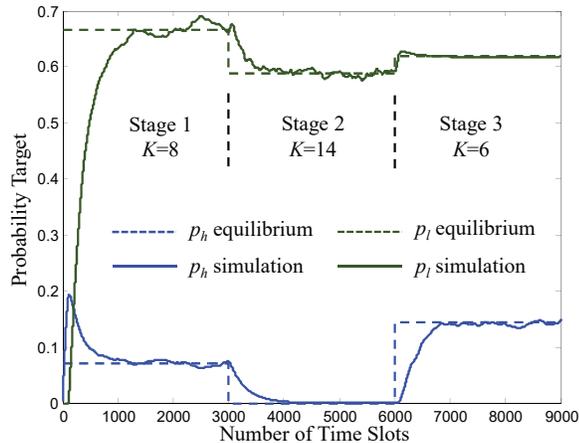}\caption{\label{MultipleOptionExampleProbabilitySimulation} Entries of the transmission probability vector target and their corresponding theoretical values.}
    \end{center}
\end{figure}
Note that the simulated throughput and probability values presented in the figures are measured using the same exponential averaging approach explained above. From Figures \ref{MultipleOptionExampleThroughputSimulation} and \ref{MultipleOptionExampleProbabilitySimulation}, we can see that the proposed MAC algorithm can indeed help users to adapt to the changes of stages and to adjust their transmission probability vectors to the new equilibrium.
\end{example}

According to the Head and Tail Condition, the system degrades to an equivalent single option system when $K\le \underline{K}$ and $K\ge \overline{K}$. It is generally expected that transmission direction vectors of the ``Head'' and the ``Tail'' regimes should be different, i.e., $\mbox{\boldmath $d$}(\underline{K})\ne \mbox{\boldmath $d$}(\overline{K})$. In Example \ref{MultipleOptionExample}, we found one virtual packet design that supports both $b_H=1.01$ in the ``Head'' regime and $b_T=1.01$ in the ``Tail'' regime. One may think that such a lucky result should be rare. Surprisingly, according to our observations, in most of the problems of interest, even though one may not always be able to get the perfect result of $b_H=b_T \approx 1$, a single virtual packet can often be designed to support close to ideal values on $J_{\epsilon_v H}$, $b_H$, $J_{\epsilon_v T}$, and $b_T$. While it is possible to extend the system design and to improve design flexibility by including the transmissions of multiple (different) virtual packets in each time slot, because performance improvement provided by such an extension is often marginal, we choose to skip the corresponding discussions in this paper.

\section{Conclusion}
We investigated distributed multiple access networking with an unknown finite number of homogeneous users. An enhanced physical-link layer interface is considered where each link layer user can be equipped with multiple transmissions. With a generally modeled link layer channel, we proposed distributed MAC algorithms to adapt the transmission schemes of the users to maximize a chosen symmetric network utility. Convergence property of the proposed MAC algorithms is proven under quite mild conditions. While there is no theoretical guarantee on the optimality of the proposed MAC algorithms, simulation results suggest that performances of the proposed MAC algorithms are often not too far from optimal.

\appendix
\subsection{Proof of Theorem \ref{MonotonicityOfqv}}
\label{MonotonictyOfqvProof}
\begin{proof}
The partial derivative of $q_v$ with respect to $p$ is given by
\begin{eqnarray}
&& \frac{\partial q_v\left( p, K \right) }{\partial p}= \sum_{j=0}^{K} \left( {K \atop j} \right) j p^{j-1} (1-p)^{K-j} C_{vj}  \nonumber \\
&& \quad -\sum_{j=0}^{K} \left( {K \atop j} \right) p^{j} (K-j) (1-p)^{K-j-1} C_{vj} \nonumber \\
&& = \sum_{j=0}^{K-1} K \left( {K-1 \atop j} \right) p^{j}  (1-p)^{K-j-1} \left( C_{v(j+1)} - C_{vj} \right) \nonumber \\
&& \le 0,
\label{Derivative_qv_p}
\end{eqnarray}
where the last inequality is due to the assumption that $C_{vj} \geq C_{v(j+1)}$ for all $j\ge 0$. Furthermore, (\ref{Derivative_qv_p}) holds with strict inequality if $K > J_{\epsilon_v}$ and $p(1-p) \ne 0$.
\end{proof}

\subsection{Proof of Theorem \ref{Monotonicity_q_vstar}}
\label{Monotonicity_q_vstarProof}
\begin{proof}
	Let us first consider the situation when $\frac{x^*}{N+b} \le p_{max}$. The derivative of $q_v^*\left( \hat{p} \right) $ with respect to $\hat{p}$ is given by
	\begin{eqnarray}
	&& \frac{d q_v^*\left( \hat{p} \right)}{d \hat{p}} = \frac{q_N\left( \hat{p} \right) - q_{N+1}\left( \hat{p} \right)}{p_N-p_{N+1}} + \frac{\hat{p} - p_{N+1}}{p_N-p_{N+1}} \frac{d q_N\left( \hat{p} \right)}{d \hat{p}} \nonumber \\
	&&  \quad + \frac{p_N - \hat{p}}{p_N - p_{N+1}} \frac{d q_{N+1} \left( \hat{p} \right)}{d \hat{p}}.
	\label{Derivative_qv*}
	\end{eqnarray}
	Write $\hat{K}=N+1-\lambda$ where $0 <\lambda \le 1$. We have
	\begin{equation}
	\hat{p}-p_{N+1}=\frac{x^*}{\hat{K}+b}-\frac{x^*}{N+1+b}=\frac{\lambda}{N+1+b} \hat{p},
	\label{Def1}
	\end{equation}
	and
	\begin{equation}
	p_{N}-\hat{p}=\frac{x^*}{N+b} - \frac{x^*}{\hat{K}+b}=\frac{1-\lambda}{N+b} \hat{p}.
	\label{Def2}
	\end{equation}
	Meanwhile, note that function $q_{N+1} \left( \hat{p} \right) $ can be decomposed as
	\begin{eqnarray}
	&& q_{N+1} \left( \hat{p} \right) = \sum_{j=0}^{N+1} \left( {N+1 \atop j} \right) \hat{p}^j (1-\hat{p})^{N+1-j} C_{vj} \nonumber \\
	&& = \hat{p} \sum_{j=0}^{N} \left( {N \atop j} \right) \hat{p}^j (1-\hat{p})^{N-j} C_{v(j+1)} \nonumber \\
	&& \quad + (1-\hat{p}) \sum_{j=0}^{N} \left( {N \atop j} \right) \hat{p}^j (1-\hat{p})^{N-j} C_{vj}.
	\label{Decomposing_q}
	\end{eqnarray}
	This leads to
	\begin{eqnarray}
	&& q_{N} \left( \hat{p} \right) - q_{N+1} \left( \hat{p} \right) = \nonumber \\
    && \quad \hat{p} \sum_{j=0}^{N} \left( {N \atop j} \right) \hat{p}^j (1-\hat{p})^{N-j} \left( C_{vj}-C_{v(j+1)} \right).
	\label{Difference}
	\end{eqnarray}
    Furthermore, the derivatives of $q_{N} \left( \hat{p} \right) $ and $q_{N+1} \left( \hat{p} \right)$ with respect to $\hat{p}$ are respectively given by
	\begin{eqnarray}
	&& \frac{d q_{N} \left( \hat{p} \right)}{d \hat{p}} = - \sum_{j=0}^{N} (N-j) \left( {N \atop j} \right) \hat{p}^j (1-\hat{p})^{N-j-1} \nonumber \\
    && \quad \times \left( C_{vj}-C_{v(j+1)} \right),
	\label{Derivative_q_N}
	\end{eqnarray}
	and
	\begin{eqnarray}
    && \frac{d q_{N+1} \left( \hat{p} \right)}{d \hat{p}} = - \sum_{j=0}^{N} (N+1) \left( {N \atop j} \right) \hat{p}^j (1-\hat{p})^{N-j} \nonumber \\
    && \quad \times \left( C_{vj}-C_{v(j+1)} \right).
	\label{Derivative_q_N+1}
	\end{eqnarray}
	Substitute (\ref{Def1}), (\ref{Def2}), (\ref{Difference}), (\ref{Derivative_q_N}), and (\ref{Derivative_q_N+1}) into (\ref{Derivative_qv*}), we get
	\begin{eqnarray}
	&& \left( p_N-p_{N+1} \right) \frac{d q_v^*\left( \hat{p} \right)}{d \hat{p}} \nonumber \\
	&& = \sum_{j=0}^{N} \left( {N \atop j} \right) \hat{p}^{j+1} (1-\hat{p})^{N-j} \left( C_{vj}-C_{v(j+1)} \right) \nonumber \\
	&&  \quad -\frac{\lambda}{N+1+b} \sum_{j=0}^{N} (N-j) \left( {N \atop j} \right) \hat{p}^{j+1}  (1-\hat{p})^{N-j-1}  \nonumber \\
    && \qquad \times \left( C_{vj}-C_{v(j+1)} \right) \nonumber \\
	&& \quad - \frac{1-\lambda}{N+b} \sum_{j=0}^{N} (N+1) \left( {N \atop j} \right) \hat{p}^{j+1} (1-\hat{p})^{N-j} \nonumber \\
    && \qquad \times \left( C_{vj}-C_{v(j+1)} \right) \nonumber \\
	&& = \sum_{j=0}^{N} \left( {N \atop j} \right) \hat{p}^{j+1} (1-\hat{p})^{N-j-1} \left( C_{vj}-C_{v(j+1)} \right) \nonumber \\
	&& \times \left( 1-\hat{p}-\frac{\lambda (N-j)}{N+1+b} -\frac{(1-\lambda) (1-\hat{p}) (N+1)}{N+b} \right) \nonumber \\
	&& =  \sum_{j=0}^{N} \left( {N \atop j} \right) \hat{p}^{j+1} (1-\hat{p})^{N-j-1} \left( C_{vj}-C_{v(j+1)} \right) \nonumber \\
	&& \quad \times \left( \frac{\lambda \left( (1-\hat{p}) (N+1+b) -N+j \right) }{N+1+b} \right. \nonumber \\
    && \quad \left. +\frac{(1-\lambda) (1-\hat{p}) (b-1)}{N+b} \right) .
	\label{Substitution}
	\end{eqnarray}
	Note that, we have the following inequalities for all $j \ge 0$.
	\begin{eqnarray}
	&& \frac{\lambda \left( (1-\hat{p}) (N+1+b)-N+j \right) }{N+1+b} \nonumber \\
	&& \ge \frac{\lambda \left( (1-p_N) (N+1+b)-N+j \right) }{N+1+b} \nonumber \\
	&& \ge \frac{\lambda \left( b-x^*+j \right) }{N+1+b}.
	\label{Inequalities}
	\end{eqnarray}
	Therefore, $\frac{d q_v^*(\hat{p})}{d \hat{p}} \ge 0$ holds if $b \ge 1$ and the following inequality is satisfied.
	\begin{equation}
	\sum_{j=0}^{N} \left( {N \atop j} \right) \hat{p}^j (1-\hat{p})^{N-1-j} \left( C_{vj}-C_{v(j+1)} \right) \left( b-x^*+j \right) \ge 0.
	\label{MonotonicityCondition}
	\end{equation}
	It is straightforward to show that (\ref{MonotonicityCondition}) holds if $b \ge x^*-\gamma_{\epsilon_v}$, where $\gamma_{\epsilon_v}$ is defined in (\ref{Gamma}).

    Furthermore, if $b > 1$ and $b > x^*-\gamma_{\epsilon_v}$ both hold with strict inequality, and considering the fact that $C_{vj} > C_{v(j+1)}$ for $j= J_{\epsilon_v}\le N$, $\frac{d q_v^*(\hat{p})}{d \hat{p}} >0 $ should be strictly positive for $\hat{p} \in (0,p_{max})$.

    Now consider the situation when $\frac{x^*}{N+b} \ge p_{max}$. If $\frac{x^*}{\hat{K}+b} \ge p_{max}$, we have $\frac{d q_v^*(\hat{p})}{d \hat{p}}= 0$. If $\frac{x^*}{\hat{K}+b} < p_{max}$ but $\frac{x^*}{N+b} \ge p_{max}$, on the other hand, we can write $\hat{K}=N+1-\lambda$ with $0 < \lambda \le N+1+b-\frac{x^*}{p_{max}}$. Consequently, (\ref{Derivative_qv*}) and (\ref{Def1}) still hold, but (\ref{Def2}) should be changed to
	\begin{equation}
	p_{N}-\hat{p}=p_{max} - \frac{x^*}{\hat{K}+b} \leq \frac{1-\lambda}{N+b} \hat{p}.
	\label{Def2_Mod}
	\end{equation}
	As a result, (\ref{Substitution}) becomes
	\begin{eqnarray}
	&& \left( p_N-p_{N+1} \right) \frac{d q_v^*\left( \hat{p} \right)}{d \hat{p}} \nonumber \\
	&& \ge \sum_{j=0}^{N} \left( {N \atop j} \right) \hat{p}^{j+1} (1-\hat{p})^{N-j-1} \left( C_{vj}-C_{v(j+1)} \right) \nonumber \\
	&& \quad \times \left( \frac{\lambda \left( (1-\hat{p}) (N+1+b) -N+j \right) }{N+1+b} \right. \nonumber \\
    && \quad \left. +\frac{(1-\lambda) (1-\hat{p}) (b-1)}{N+b} \right).
	\label{SubstitutionModification}
	\end{eqnarray}
	By following the rest of the derivations, it can be seen that conclusion of the theorem still holds.	
\end{proof}

\subsection{Proof of Theorem \ref{ConvergenceMACAlgorithm}}
\label{ProofConvergenceMACAlgorithm}
\begin{proof}
    We first show that the associated ODE given in (\ref{AssociatedSystemODE}) has a unique equilibrium at $\mbox{\boldmath $p$}^* = \min \{p_{\max}, \frac{x^*}{K+b}\} \mbox{\boldmath $1$}$. Because $b > \max \{1, x^*-\gamma_{\epsilon_v}\}$, theoretical channel contention measure $q_v^*(\hat{p})$ is strictly increasing in $\hat{p}$ for $\hat{p} \in (0,p_{\max})$. Given user number $K$, $q_v(\hat{p}, K)$ is non-increasing in $\hat{p}$. Therefore, if $K\ge J_{\epsilon_v}$, then $\hat{p}=p^*=\frac{x^*}{K+b}$ is the only solution to $q_v(\hat{p}, K)=q_v^*(\hat{p})$. If $K<J_{\epsilon_v}$, on the other hand, we must have $q_v(\hat{p}, K)>q_v^*(\hat{p})$ for all $\hat{p}\in [0, p_{\max})$. This implies that $\mbox{\boldmath $p$}^* = \min \{p_{\max}, \frac{x^*}{K+b}\} \mbox{\boldmath $1$}$ is the only equilibrium of the system.

    Second, we show that there exists a constant $\epsilon>0$, such that $\frac{d q_v^*(\hat{p})}{d \hat{p}} \ge \epsilon >0$ for all $\hat{p} < p_{\max}$. Note that $\hat{p} < p_{\max}$ implies $\hat{K} > J_{\epsilon_v}$. Therefore, according to (\ref{Substitution}) and (\ref{Inequalities}), we have
    \begin{eqnarray}
	&& \frac{d q_v^*(\hat{p})}{d \hat{p}} \ge \frac{\hat{p}}{p_N - p_{N+1}} \left( {N \atop J_{\epsilon_v}} \right) \hat{p}^{J_{\epsilon_v}} (1-\hat{p})^{N-J_{\epsilon_v}-1} \nonumber \\
	&&  \quad \times (C_{vJ_{\epsilon_v}} - C_{v(J_{\epsilon_v}+1)}) \nonumber \\
	&& \times \left( \frac{\lambda (b-x^*+J_{\epsilon_v})}{N+1+b} + \frac{(1-\lambda)(1-\hat{p})(b-1)}{N+b} \right).
	\label{Bound_Derivative}
	\end{eqnarray}
    Because the right hand side of (\ref{Bound_Derivative}) has a positive limit as $\hat{p}$ is taken to zero, we can find two small positive constants $\epsilon_0, \epsilon_1>0$, such that $\frac{d q_v^*(\hat{p})}{d \hat{p}} \ge \epsilon_0 >0$ for all $\hat{p}\le \epsilon_1$. On the other hand, when $\epsilon_1 \le \hat{p}<p_{\max}$, because $b > \max \{1, x^*-\gamma_{\epsilon_v}\}$ holds with strict inequality, we can find a small positive constant $\epsilon_2>0$, such that the right hand side of (\ref{Bound_Derivative}) is no less than $\epsilon_2$. Consequently, choose $\epsilon=\min\{\epsilon_0, \epsilon_2\}$, we have $\frac{d q_v^*(\hat{p})}{d \hat{p}} \ge \epsilon >0$ for all $\hat{p} < p_{\max}$.

    Third, let $q_v^{*-1}(.)$ denote the inverse function of $q_v^*(\hat{p})$. Then, for every transmission probability vector $\mbox{\boldmath $p$}$, the target transmission probability is obtained by
	\begin{equation}
	\hat{p} = q_v^{*-1} (q_v) = q_v^{*-1} (q_v(\mbox{\boldmath $p$},K)).
	\end{equation}
    Because $\frac{d q_v^*(\hat{p})}{d \hat{p}} \ge \epsilon >0$ for all $\hat{p} < p_{\max}$, one can find a constant $K_{l_1}$ such that
	\begin{equation}
	| \hat{p}_1 - \hat{p}_2 | \leq K_{l_1} | q_{v_1} - q_{v_2} |,
	\label{Inequality1}
	\end{equation}
    for all $\hat{p}_1 = q_v^{*-1}(q_{v_1})$ and $\hat{p}_2 = q_v^{*-1}(q_{v_2})$. Since $q_v = q_v(\mbox{\boldmath $p$}, K)$ is Lipschitz continuous in $\mbox{\boldmath $p$}$, there exists a constant $K_{l_2}$ to satisfy
	\begin{equation}
	| q_{v_1} - q_{v_2} | \leq K_{l_2} \| \mbox{\boldmath $p$}_1 - \mbox{\boldmath $p$}_2 \|,
	\label{Inequality2}
	\end{equation}
    for all $q_{v_1} = q_v(\mbox{\boldmath $p$}_1,K)$ and $q_{v_2} = q_v(\mbox{\boldmath $p$}_2,K)$. By combining (\ref{Inequality1}) and (\ref{Inequality2}), we have
	\begin{equation}
	| \hat{p}_1 - \hat{p}_2 | \leq K_{l_1} K_{l_2} \| \mbox{\boldmath $p$}_1 - \mbox{\boldmath $p$}_2 \|,
	\label{IneLipschitz}
	\end{equation}
    for all $\hat{p}_1 = q_v^{*-1}(q_v(\mbox{\boldmath $p$}_1,K))$ and $\hat{p}_2 = q_v^{*-1}(q_v(\mbox{\boldmath $p$}_2,K))$. Therefore, $\hat{p}(\mbox{\boldmath $p$})$ is Lipschitz continuous and satisfies Condition \ref{LipschitzAssumption}.
	
    Finally, when the system is noisy, the receiver can choose to measure $q_v$ over an extended number of time slots, or equivalently, to increase the value of $Q$ introduced in Step \ref{SingleOptionStep2} of the proposed MAC algorithm. If users maintain their transmission probabilities during the $Q$ time slots, it is often the case that the potential measurement bias in the system can be reduced arbitrarily close to zero. Therefore, the Mean and Bias Condition \ref{MeanBiasAssumption} is also satisfied.

    Consequently, convergence of proposed distributed MAC algorithm is guaranteed by Theorems \ref{StochasticProbabilityOneConvergence} and \ref{StochasticWeakConvergence}.
\end{proof}

\subsection{Proof of Theorem \ref{TheoremOfConvergence2}}
\label{ProofofTheoremOfConvergence2}
\begin{proof}
    We first show that the associated ODE given in (\ref{AssociatedSystemODE}) should have a unique equilibrium at $\mbox{\boldmath $P$}^* = \mbox{\boldmath $1$} \otimes \mbox{\boldmath $p$}(K)$. According to Condition \ref{MonotoncityGradientCondition} and Theorem \ref{Monotonicity_q_vstar}, $q_v^*(\hat{K})$ is strictly decreasing in $\hat{K}$ for $\hat{K} \ge J_{\epsilon_v}(\mbox{\boldmath $d$}(\underline{K}))$. Because we set $\hat{K}=J_{\epsilon_v}(\mbox{\boldmath $d$}(\underline{K}))$ if $q_v > q_v^*( J_{\epsilon_v}(\mbox{\boldmath $d$}(\underline{K})))$, any equilibrium of the ODE must take the form of $\mbox{\boldmath $P$}^* = \mbox{\boldmath $1$} \otimes \mbox{\boldmath $p$}(\hat{K})$ for a $\hat{K}\ge J_{\epsilon_v}(\mbox{\boldmath $d$}(\underline{K}))$.

    Assume that the actual user number satisfies $\underline{K}\le K \le \overline{K}$. With users setting their transmission probability vectors at $\mbox{\boldmath $p$}(\hat{K})$, due to Condition \ref{MonotoncityGradientCondition}, if $K>\hat{K}$ and $\hat{K}$ is an integer, we must have
    \begin{equation}
	q_v(\mbox{\boldmath $p$}(\hat{K}), K) < q_v(\mbox{\boldmath $p$}(\hat{K}), \hat{K})=q_v^*(\hat{K}).
	\label{EquInequality1}
	\end{equation}
	If $K>\hat{K}$ and $\hat{K}$ is not an integer, we have
    \begin{eqnarray}
	&& q_v(\mbox{\boldmath $p$}(\hat{K}), K) < q_v(\mbox{\boldmath $p$}(\hat{K}), \lfloor \hat{K} \rfloor), \nonumber \\
	&& q_v(\mbox{\boldmath $p$}(\hat{K}), K) \le q_v(\mbox{\boldmath $p$}(\hat{K}), \lfloor \hat{K} \rfloor+1),
	\label{EquInequality2}
	\end{eqnarray}
	which implies that
    \begin{equation}
    q_v(\mbox{\boldmath $p$}(\hat{K}), K) < q_v^*(\hat{K}).
    \end{equation}
    On the other hand, if $K< \hat{K} $ and $\hat{K}$ is an integer, we must have
    \begin{equation}
	q_v(\mbox{\boldmath $p$}(\hat{K}), K) > q_v(\mbox{\boldmath $p$}(\hat{K}), \hat{K})=q_v^*(\hat{K}).
	\label{EquInequality3}
	\end{equation}
	If $K< \hat{K} $ and $\hat{K}$ is not an integer, we have
    \begin{eqnarray}
	&& q_v(\mbox{\boldmath $p$}(\hat{K}), K) > q_v(\mbox{\boldmath $p$}(\hat{K}), \lfloor \hat{K} \rfloor+1), \nonumber \\
	&& q_v(\mbox{\boldmath $p$}(\hat{K}), K) \ge  q_v(\mbox{\boldmath $p$}(\hat{K}), \lfloor \hat{K} \rfloor),
	\label{EquInequality4}
	\end{eqnarray}
    which also implies that
    \begin{equation}
    q_v(\mbox{\boldmath $p$}(\hat{K}), K) > q_v^*(\hat{K}).
    \end{equation}
    Consequently, (\ref{EstimatingUserNumber}) must have a unique solution at $\hat{K}=K$. When $K\le \underline{K}$ or $K\ge \overline{K}$, on the other hand, uniqueness of the solution to (\ref{EstimatingUserNumber}) can be shown by following the proof of Theorem \ref{ConvergenceMACAlgorithm}.

    Second, by Condition \ref{MonotoncityGradientCondition}, for $\underline{K} \le \hat{K} \le \overline{K}$, $\mbox{\boldmath $p$}(\hat{K})$ is Lipschitz continuous in $\hat{K}$ and $q_v^*(\hat{K})$ satisfies (\ref{Lipschitzqvstar}). Combined with the Head and Tail Condition \ref{HeadTailAssumption} and the fact that $p(\hat{K})$ is designed for $\hat{K} \le \underline{K}$ and $\hat{K} \ge \overline{K}$ according to the guideline presented in Section \ref{SingleOption}, we conclude that $\mbox{\boldmath $p$}(q_v)$ is Lipschitz continuous in $q_v$. Because $q_v(\mbox{\boldmath $P$}, K)$ is also Lipschitz continuous in $\mbox{\boldmath $P$}$, $\mbox{\boldmath $p$}(q_v(\mbox{\boldmath $P$}, K))$ must be Lipschitz continuous in $\mbox{\boldmath $P$}$, and hence Condition \ref{LipschitzAssumption} is satisfied.

    Finally, the Mean and Bias Condition \ref{MeanBiasAssumption} is satisfied because, by assumption, the receiver can increase $Q$ in Step \ref{Step2} of the proposed MAC algorithm to reduce the potential measurement bias arbitrarily close to zero.
\end{proof}

\subsection{Proof of Theorem \ref{PinPointsTheorem}}
\label{ProofOfPinPointsTheorem}
\begin{proof}
    Because Items \ref{GradientConditionItem2}, \ref{GradientConditionItem3}, and \ref{GradientConditionItem4} in Condition \ref{MonotoncityGradientCondition} hold by assumption, we only need to prove Item \ref{GradientConditionItem1} in Condition \ref{MonotoncityGradientCondition}. That is, with the Interpolation Approach, $\mbox{\boldmath $p$}(\hat{K})$ should be Lipschitz continuous in $\hat{K}$. For the sake of simple notation, we use $\frac{d p(\hat{K})}{d \hat{K}}$ to represent the derivative of $p(\hat{K})$ with respect to $\hat{K}$ if $p(\hat{K})$ is differentiable. If $p(\hat{K})$ is only continuous but not differentiable at $\hat{K}$, then $\frac{d p(\hat{K})}{d \hat{K}}$ represents one or an arbitrary subderivative of $p(\hat{K})$. If $p(\hat{K})$ is not continuous at $\hat{K}$, then $\frac{d p(\hat{K})}{d \hat{K}}$ should take the values of $\pm \infty$. Note that the adoption of such a notation does not imply a continuity assumption on $p(\hat{K})$.

    Let $i\in \{1, \dots, L\}$ and $0\le \lambda <1$ be chosen arbitrarily. Let $\hat{K}=\hat{K}_{i\lambda}$. To simplify the discussion, we assume that the neighboring two pinpoints satisfy $\hat{K}_{i+1} = \hat{K}_i + 1$, i.e., they take neighboring integer values\footnote{The proof can be easily extended to the case when this assumption does not hold.}. Write $\hat{K} = \hat{K}_{i\lambda} = (1-\lambda) \hat{K}_i + \lambda \hat{K}_{i+1}$ as a function of $\lambda$, we have $\frac{d p(\hat{K})}{d \hat{K}} = \frac{d p(\lambda)}{d \lambda}$.
	
    To bound $\frac{d p(\lambda)}{d \lambda}$, we consider two different expressions of $q_v^*(\hat{K})=q_v^*(\lambda)$. The first expression is
    \begin{equation}
    q^*_v(\lambda) = (1-\lambda) q^*_v(\hat{K}_i) + \lambda q^*_v(\hat{K}_{i+1}).
    \label{qvstarfirstexpression}
    \end{equation}
    Take derivative with respect to $\lambda$, we get $\frac{d q_v^*(\lambda)}{d\lambda} = q_v^*(\hat{K}_{i+1})- q_v^*(\hat{K}_i)$. Because both $q_v^*(\hat{K}_{i+1})$ and $q_v^*(\hat{K}_i)$ are bounded, there exists a positive constant $\overline{\Delta}_1>0$ such that
	\begin{equation}
	\left| \frac{d q^*_v(\lambda)}{d \lambda} \right|  \le \overline{\Delta}_1.
	\label{BoundDerivative_q*_lambda}
	\end{equation}

    On the other hand, consider the second expression of $q_v^*(\hat{K})=q_v^*(\lambda)$ given below.
    \begin{equation}
    q^*_v(\lambda, p_{i\lambda} \mbox{\boldmath $d$}_{i\lambda}) = (1-\lambda) q_v(p_{i\lambda} \mbox{\boldmath $d$}_{i\lambda}, \hat{K}_i) + \lambda q_v(p_{i\lambda} \mbox{\boldmath $d$}_{i\lambda}, \hat{K}_{i+1}).
    \label{qvstarsecondexpression}
    \end{equation}
    Taking derivative with respect to $\lambda$ results in
	\begin{eqnarray}
	&& \frac{d q^*_v(\lambda, p_{i\lambda} \mbox{\boldmath $d$}_{i\lambda})}{d \lambda} = \frac{\partial q^*_v(\lambda, p_{i\lambda} \mbox{\boldmath $d$}_{i\lambda})}{\partial \lambda} \nonumber \\
    && \quad +\left[ \frac{\partial q^*_v(\lambda, p_{i\lambda} \mbox{\boldmath $d$}_{i\lambda})}{\partial \mbox{\boldmath $d$}_{i \lambda}} \right]^T \frac{d \mbox{\boldmath $d$}_{i \lambda}}{d \lambda} + \frac{\partial q^*_v(\lambda, p_{i\lambda} \mbox{\boldmath $d$}_{i\lambda}) }{\partial p_{i \lambda}} \frac{d p_{i \lambda}}{ d \lambda}. \nonumber \\
	\label{AnotherDerivative_q*_lambda}
	\end{eqnarray}
	Now we consider the terms on the right hand side of (\ref{AnotherDerivative_q*_lambda}) separately.
	\begin{equation}
	\frac{\partial q^*_v(\lambda, p_{i\lambda} \mbox{\boldmath $d$}_{i\lambda})}{\partial \lambda} = q_v(p_{i \lambda} \mbox{\boldmath $d$}_{i \lambda}, \hat{K}_{i+1}) - q_v(p_{i \lambda} \mbox{\boldmath $d$}_{i \lambda}, \hat{K}_{i}).
	\label{PartialDerivative_lambda}
	\end{equation}
	Because both two terms on the right hand side of (\ref{PartialDerivative_lambda}) are bounded, there exists a constant $\overline{\Delta}_2>0$ to satisfy
	\begin{equation}
	\left| \frac{\partial q^*_v(\lambda, p_{i\lambda} \mbox{\boldmath $d$}_{i\lambda})}{\partial \lambda} \right|  \le \overline{\Delta}_2.
	\label{BoundDerivative_lambda}
	\end{equation}
	By following a derivation similar to (\ref{Derivative_qv_p}), it can be verified that there exists a constant $\overline{\Delta}_3>0$ such that
	\begin{equation}
	\left| \left[ \frac{\partial q^*_v(\lambda, p_{i\lambda} \mbox{\boldmath $d$}_{i\lambda})}{\partial \mbox{\boldmath $d$}_{i \lambda}} \right]^T \frac{d \mbox{\boldmath $d$}_{i \lambda}}{d \lambda} \right|  \leq \overline{\Delta}_3.
	\label{BoundDerivative_directio}
	\end{equation}
    Because $\hat{K} > \underline{K} \geq J_{\epsilon_v}(\mbox{\boldmath $d$}(\underline{K}))$ and $\underline{p} \leq p(\hat{K})=p_{i \lambda} \leq \overline{p}$, from the derivation similar to (\ref{Derivative_qv_p}), we can see there exists a positive constant $\underline{\Delta}_1>0$ such that
	\begin{equation}
	\left| \frac{\partial q^*_v(\lambda, p_{i\lambda} \mbox{\boldmath $d$}_{i\lambda})}{\partial p_{i\lambda}} \right|  \geq \underline{\Delta}_1.
	\label{BoundDerivative_p}
	\end{equation}

    Because the two expressions of $q_v^*(\hat{K})$ given in (\ref{qvstarfirstexpression}) and (\ref{qvstarsecondexpression}) must equal each other, by combining (\ref{BoundDerivative_q*_lambda}), (\ref{AnotherDerivative_q*_lambda}), (\ref{BoundDerivative_lambda}), (\ref{BoundDerivative_directio}), and (\ref{BoundDerivative_p}), we conclude that there exists a positive constant $K_g>0$, such that $\left\| \frac{d \mbox{\boldmath $p$}(\hat{K})}{d \hat{K}} \right\| \leq K_g $. With the extended definition of $\frac{d \mbox{\boldmath $p$}(\hat{K})}{d \hat{K}}$, as explained at the beginning of the proof, $\left\| \frac{d \mbox{\boldmath $p$}(\hat{K})}{d \hat{K}} \right\| \leq K_g $ means that $\mbox{\boldmath $p$}(\hat{K})$ is Lipschitz continuous in $\hat{K}$.
\end{proof}





%




\end{document}